\documentclass[prd,aps,floats,floatfix,nofootinbib]{revtex4}
\usepackage{amsmath}
\usepackage{amsfonts}
\usepackage{graphicx}
\usepackage{dcolumn}
\usepackage{bm}
\usepackage[pdftex]{color}
\usepackage{slashed}
\newcommand{\beq}{\begin{equation}}
\newcommand{\eeq}{\end{equation}}
\def\fsl#1{\setbox0=\hbox{$#1$}           
   \dimen0=\wd0                                 
   \setbox1=\hbox{/} \dimen1=\wd1               
   \ifdim\dimen0>\dimen1                        
      \rlap{\hbox to \dimen0{\hfil/\hfil}}      
      #1                                        
   \else                                        
      \rlap{\hbox to \dimen1{\hfil$#1$\hfil}}   
      /                                         
   \fi}                                         %

\begin{document}
\title{\Large Patterns of  Custodial Isospin Violation from a Composite Top}

\author{R. Sekhar Chivukula$^{a}$}
\email[]{sekhar@msu.edu}
\author{Roshan Foadi$^{b}$}
\email[]{rfoadi@ulb.ac.be}
\author{Elizabeth H. Simmons$^a$}
\email[]{esimmons@msu.edu}
\affiliation{$^a$Department of Physics and Astronomy,
Michigan State University, East Lansing, MI 48824, USA\\
$^b$ Service de Physique Th\'eorique, Universit\'e Libre de Bruxelles, \\
Campus de la Plaine CP225, Bd du Triomphe, 1050 Brussels, Belgium
}


\begin{abstract}
 
 In this paper we consider the effects of top quark compositeness on the 
 electroweak parameters  $\hat{T}$ and $\hat{S}$ and 
 the $Z b_L \bar{b}_L$ coupling. We do so by using an effective field theory analysis to identify several promising patterns of mixing between SM-like and vector fermions, and then analyzing simple extensions of the
 Standard Model that realize those patterns. These models illustrate four ways in which 
 an extended $O(4) $ symmetry, which controls the size of radiative  
 corrections to the observables discussed, may be broken. These models may 
 also be viewed as highly-deconstructed versions of five-dimensional gauge theories 
 dual to various strongly-interacting composite Higgs theories. We comment on how our results relate to
 extra-dimensional models previously considered, and we demonstrate 
 that one pattern of $O(4) $ breaking is phenomenologically favored.
 
\end{abstract}

\maketitle

\section{Introduction}

New strong interactions are a natural possibility for the dynamics underlying electroweak
symmetry breaking. In analogy with the AdS/CFT 
correspondence \cite{Maldacena:1997re,Gubser:1998bc,Witten:1998qj}, such dynamics are expected to have a weakly-coupled dual description in 
terms of a compactified five-dimensional gauge theory. In the dual five-dimensional description,
the ordinary electroweak gauge-bosons are understood as the lightest ``Kaluza-Klein" resonances
of the five-dimensional gauge theory, whose light masses arise either through the boundary 
conditions imposed on the five-dimensional gauge fields, as in the case of Higgsless models~\cite{Csaki:2003dt}, 
or through the vacuum expectation value of a composite scalar Higgs~\cite{ArkaniHamed:2001nc,Contino:2003ve}.

In both Higgsless \cite{Cacciapaglia:2004rb,Foadi:2004ps,Chivukula:2005xm} and composite Higgs models \cite{Agashe:2003zs,Agashe:2004rs}, the observed mass-eigenstate fermions result from mixing between two kinds of gauge-eigenstate fermions: a set with quantum numbers resembling those of Standard Model fermions and a new set of vector fermions \cite{Contino:2006nn,Pomarol:2008bh,Kaplan:1991dc}.  The first set are elementary (non-composite) fermion fields that are only weakly coupled
to the new strong dynamics, and they correspond to ``brane-localized" states that are largely confined to the ultraviolet boundary 
of the compactified space. In contrast, the new vector fermions are Kaluza-Klein resonances arising from the ``bulk fields" of the compactified five-dimensional theory, and in the dual four-dimensional theory they correspond to a tower of composite states arising from the underlying strong dynamics.  The fermions observed in experiment correspond to the lighter mass eigenstates resulting from this mixing, and they are mostly composed of elementary fields with a smaller admixture of vector states; we will refer to these as "ordinary fermions".  The heavier partner mass eigenstates, which have not yet been observed, are predominantly composed of the composite, vectorial states; we will refer to these as ``heavy vector fermions".  Their mass scale $M$ is typically in the $10^2-10^3$ GeV range.

\begin{figure}[!t]
\includegraphics[width=10cm,height=4cm]{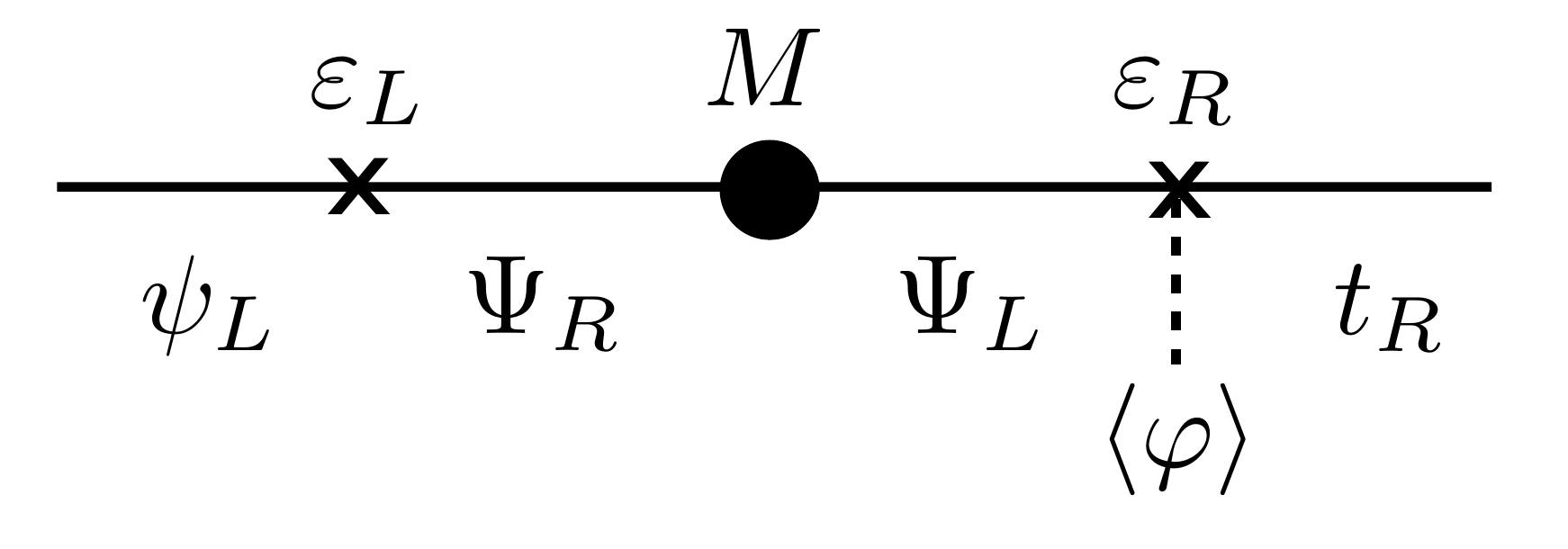}
\caption{Schematic self-energy diagram for the top quark when mixing occurs between fundamental SM-like gauge eigenstates $(\psi_L, t_R)$ and composite vector fermions $\Psi$, and a Yukawa coupling between $\Psi_L$ and $t_R$ is mediated by scalar $\varphi$.  As discussed in the text, the mass of the lighter mass eigenstate will depend on the Dirac mass $M$ of the vector fermions, and the mixing factors $(\varepsilon_L, \varepsilon_R)$.}
\label{fig:mass-fig}
\end{figure}

The composition of the lighter mass eigenstates will affect their properties.   Consider, for instance, an elementary brane-localized fermion $t$, whose left-handed component is a member of the weak doublet $\psi_L$, and whose right-handed component is an electroweak singlet.  We will denote by $\varepsilon_L$ and $\varepsilon_R$, respectively, 
the degree of mixing of the weak doublet $\psi_L$ and weak singlet $t_R$ fields with the new vector fermions $\Psi$  bearing the same Lorentz and electroweak quantum numbers.    Then, as suggested by Fig. \ref{fig:mass-fig}, the  mass of the lighter ``ordinary fermion" mass eigenstate resulting from the mixing will be of order $m_t\sim \varepsilon_L \varepsilon_R M$.  It is then clear that in order for the light flavors of the ordinary fermions to receive their appropriate masses  $\varepsilon_L$ and/or $\varepsilon_R$ must be quite small, whereas for the top quark neither $\varepsilon_L$ nor $\varepsilon_R$ can be too small if $m_t$ is to have its observed value. Hence, from the four-dimensional point of view, the light ordinary fermions will be essentially elementary, while the top-quark must be substantially composite.

In these models, the hierarchy of ordinary fermion masses is then transferred to 
a hierarchy of the values of the mixing factors $\varepsilon$.   In models with a warped extra-dimension, a possible 
explanation for such a hierarchy, results from the combination of the exponential factor in the AdS$_5$ metric, along with different fermion profiles in the bulk ~\cite{Grossman:1999ra,Gherghetta:2000qt,Kaplan:2001ga,Csaki:2003sh}. We will assume that the strong electroweak breaking dynamics incorporates either minimal 
\cite{Chivukula:1987py,D'Ambrosio:2002ex} or next-to-minimal \cite{Agashe:2004cp} flavor violation, and therefore that non-SM
contributions of these strong dynamics to flavor-changing neutral currents are sufficiently suppressed to avoid conflict with experiment.

Fermion compositeness can yield significant corrections to low-energy observables; some are beneficial and others are problematic. In
Higgsless models, for example, the presence of vector fermions with SM quantum numbers can cancel  
contributions to the $S$ parameter arising from the extended gauge sector \cite{Cacciapaglia:2004rb,Foadi:2004ps,Chivukula:2005xm}.
Additional effects are expected to arise predominantly from the top sector, where the mixing factors $\epsilon$ are largest. Two electroweak quantities are particularly sensitive to effects in the top-sector: the $Z b_L \bar{b}_L$ coupling, $g_{Lb}$, and the
deviation in the ratio of the $W$- and $Z$-boson masses from that
predicted in the SM, also known as
the electroweak $\hat{T}$ parameter \cite{Peskin:1990zt,Altarelli:1990zd,Barbieri:2004qk,Chivukula:2004af}.
Potential contributions to both parameters can be understood in terms of the (approximate) global
symmetry structure of the new strong dynamics. Just as in the SM, contributions to $\hat{T}$ can
be suppressed if the symmetry breaking sector has an approximate $SU(2)_L \times SU(2)_R$
symmetry which, via electroweak symmetry breaking, breaks to a diagonal custodial $SU(2)_c$
symmetry \cite{Weinstein:1973gj,Sikivie:1980hm}. Furthermore, as shown by Agashe {\em et. al.} \cite{Agashe:2006at}, corrections to the $Z b_L \bar{b}_L$
coupling can be suppressed if custodial symmetry is extended to include a left-right
parity symmetry $P_{LR}$  -- whose action consists in exchanging $SU(2)_L$ and $SU(2)_R$ 
charges  -- and making $b_L$ an eigenstate of $P_{LR}$. In this case, the required
overall symmetry structure is $O(4) \sim SU(2)_L \times SU(2)_R \times P_{LR} \to SU(2)_c \times P_{LR}\sim O(3)_c $.

Due to the mass splitting between the top- and bottom-quark, neither the conventional $SU(2)_c$ nor the extended $O(3)_c$ symmetry can be exact, even in the limit of zero hypercharge.
$SU(2)_c$ requires that the top and bottom quarks form a doublet, and hence $m_b=m_t$
in the symmetry limit.  $O(3)_c$ symmetry requires that the top and bottom quark
masses both equal that of an additional exotic quark of charge +5/3, with which they form a triplet
 \cite{Agashe:2006at,SekharChivukula:2009if}. In the SM, the dimension-four
top-quark Yukawa coupling $y_t$ breaks $SU(2)_c$ weakly, and is
responsible for transmitting electroweak symmetry breaking to the top-quark
sector. Therefore, the leading SM corrections to both $\hat{T}$ and $g_{Lb}$  are proportional to 
$y^2_t/16\pi^2$.  In models in which the top-quark is composite, {\it i.e.} has a large vector fermion component, the vector fermions can be members of custodial singlets, doublets, or triplets. However, unlike in the SM and
simple generalizations thereof \cite{Einhorn:1981cy}, 
both positive and negative 
corrections to $\hat{T}$ and $g_{Lb}$ can be generated.  This is both dangerous and interesting.
It is dangerous because the SM predictions for $\hat{T}$ and $g_{Lb}$ are in agreement with experiment. It is interesting 
because the agreement is not perfect: the SM  prediction for $g_{Lb}$ is about $2\sigma$ below the measured value 
(with weak dependence on the Higgs mass) \cite{:2005ema}, whereas the SM prediction for $\hat{T}$ is in full agreement with experiment for a light Higgs ($m_H=115$ GeV) but is $2\sigma$ below for a heavy Higgs ($m_H=800$ GeV)\footnote{Of course a heavy Higgs is incompatible with the SM, unless accompanied by additional new physics.} \cite{Barbieri:2004qk}. Moreover, the measured left-handed and right-handed $Z b \bar{b}$ couplings have a strong and positive correlation, and the SM predictions are both approximately $2\sigma$ below their expectation values. 

In this paper we consider the effect of top compositeness on $\hat{T}$ and $g_{Lb}$, and discuss
the patterns of top compositeness that can yield phenomenologically viable models.  To begin, we examine the low-energy operators induced \cite{Pomarol:2008bh,SekharChivukula:2009if} when the heavy vector fermion states are ``integrated out" at tree-level, and correlate the phenomenological properties of these operators with the custodial quantum numbers of these fermions.  In sections III and IV, we construct simple models that illustrate these effects by extending the SM via the addition of one  weak-charged vector fermion multiplet mixing with the left-handed elementary top-bottom doublet, and/or one weak-singlet fector fermion mixing with the elementary right-handed top. These models may be considered as highly deconstructed versions~\cite{Contino:2006nn,Pomarol:2008bh} of full five-dimensional duals~\cite{Carena:2003fx,Carena:2004zn,Carena:2006bn,Carena:2007ua} to various underlying strongly-interacting composite Higgs theories: deconstructions that de-scope the theory to include only one non-standard Kaluza-Klein fermion level.  
Taken together, the simple models discussed illustrate the various ways in which the third-generation fermion masses arise in any strongly-interacting composite Higgs theory with a weakly coupled five-dimensional dual. Finally, in
section~\ref{sec:conclusions} we compare our results to previous calculations in five-dimensional models, note what happens if the right-handed top is mixes with a triplet state, and present our conclusions.

\section{Patterns of Custodial Isospin Violation}

\label{sec:operators}

In this section we use an effective field theory analysis of top-quark compositeness to understand how integrating out heavy vector fermion states with different quantum numbers correlate with likely phenomenological effects on $\hat{T}$ and $\delta g_{Lb}$.  We will consider, in turn, the effects of mixing of new vector fermions with each of the third-generation quark states: the top-bottom doublet, the right-handed top, and the right-handed bottom quark, in situations with and without an overall custodial symmetry.  We will catalog\footnote{A related analysis in a different language has been carried out in \cite{Pomarol:2008bh}, which noted that
in cases in which the top-quark mass arises through mixing with (composite) fermions, 
the leading low-energy effects may be summarized through the operators 
\begin{equation}
\frac{i \tilde{c}_R y^2_t}{M^2} (\varphi^\dagger D_\mu \varphi)(\bar{t}_R \gamma^\mu t_R) + 
\frac{i\tilde{c}^{(1)}_L y^2_t} {M^2} (\varphi^\dagger D_\mu \varphi)(\bar{t}_L \gamma^\mu t_L) +
\frac{i\tilde{c}^{(3)}_L y^2_t} {2 M^2} (\varphi^\dagger \sigma^a  D_\mu \varphi)(\bar{t}_L \sigma^a \gamma^\mu t_L)~,
\end{equation}
where $y_t$ is a Yukawa coupling and $M$ is the mass scale of the heavy vector fermions. The leading effects on
both $\hat{T}$ and $\delta g_{Lb}$ may then be computed in the low-energy effective
field theory \cite{Pomarol:2008bh,SekharChivukula:2009if}. Each of the operators we discuss can be recast in this language for particular choices of the coefficients $\tilde{c}_{L,R}$.}. the operators that arise from integrating out the heavy vector fermions and identify which are most likely to have significant effects.  Then, in the following sections of the paper, we will explore those operators more fully by constructing models whose low-energy effective theories give rise to them. 

\begin{figure}[!t]
\includegraphics[width=8cm,height=4.5cm]{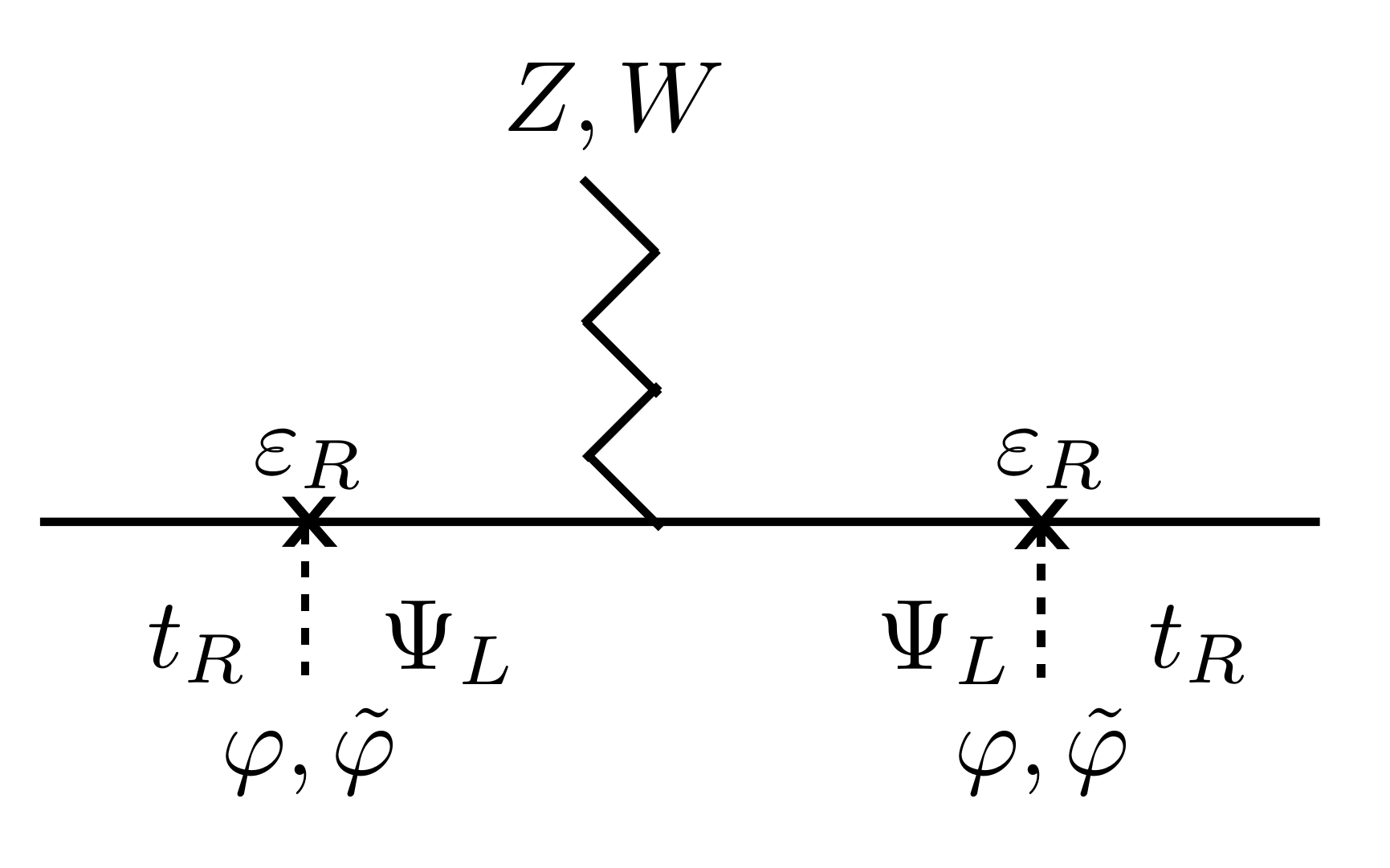}
\caption{A heavy vector-fermion $\Psi$ mixing with the left-handed top quark can give rise to new effects on the $Z\bar{t}_R t_R$ vertex in the low-energy effective theory.}
\label{fig:Ztt-fig}
\end{figure}

To set the stage, let us review a scenario from \cite{SekharChivukula:2009if} where the Lagrangian terms that include the heavy vector fermion $\Psi$ take the form
\begin{equation}
{\cal L}_{\Psi} = i \bar{\Psi} D\!\!\!\!/ \,\Psi - M \bar{\Psi}\Psi - \lambda_t \bar{\Psi}_L \varphi\, t_R + h.c.
\label{eq:simplelag}
\end{equation}
where $\varphi$ is a scalar coupling $\Psi_L$ to $t_R$. Requiring the variation of ${\cal L}_{\Psi}$ with respect to $\bar{\Psi}_{L,R}$ to vanish yields the equations of motion
\begin{align}
i D\!\!\!\!/ \,\Psi_L - M \Psi_R  = \lambda_t \varphi\, t_R \qquad\qquad
i D\!\!\!\!/ \,\Psi_R - M \Psi_L = 0~,
\end{align}
which we may solve iteratively in $1/M$. Doing so, we find
\begin{align}
\Psi_R  = -\,\frac{\lambda_t}{M}\varphi\, t_R+{\cal O}\left(\frac{ (i D\!\!\!\! /\,)^2 \varphi\, t_R}{M^3}\right) \qquad\qquad
\Psi_L  = - \frac{\lambda_t}{M^2} i D\!\!\!\!/ \,(\varphi\, t_R)+{\cal O}\left(\frac{ (i D\!\!\!\! /\,)^3 \varphi\, t_R}{M^4}\right)~.
\end{align}
Plugging these expressions into Eq. (\ref{eq:simplelag}), we obtain the following non-SM operator in the low-energy effective theory
\begin{equation}
{\cal L}_{eff} = \frac{\lambda^2_t}{M^2} \bar{t}_R \varphi^\dagger i D\!\!\!\!/ \,(\varphi\, t_R) +  \ldots~,
\label{eq:newterm}
\end{equation}
where subsequent terms are suppressed by higher powers of $1/M^2$.  In unitary gauge this term gives rise to an ``anomalous" coupling
of the $Z$-boson to top-quarks, 
\begin{equation}
\frac{\lambda_t^2}{M^2} \bar{t}_R \varphi^\dagger (i D\!\!\!\! /\,) \varphi\, t_R\ \ \longrightarrow \ \ \frac{e \lambda_t^2 v^2}{4 s_w c_w  M^2}\  \bar{t}_R Z\!\!\!\! /\ \, t_R
\end{equation}
as sketched in Fig. \ref{fig:Ztt-fig}.  Since the operator's effects on observables are governed by $\lambda_t$, they are related to the size of the top-quark mass and therefore are potentially large.

In fact, we may systematically catalog the operators that can arise from integrating out each of the possible kinds of heavy vector fermions and think about their characteristic phenomenologies  When a left-handed state mixing with the left-handed top-bottom doublet has been integrated out, either or both of the following operators may result
\begin{eqnarray}
{\cal O}_1 \equiv (\bar{t}_R\tilde{\varphi}^\dagger) i \slashed{D} (\tilde{\varphi} t_R) \label{eq:smtl-op}\\
{\cal O}_2 \equiv  (\bar{t}_R { \varphi}^\dagger) i \slashed{D} ({ \varphi} t_R) \label{eq:extl-op}
\end{eqnarray}
where $\widetilde{\varphi}\equiv i\sigma^2\varphi^\ast$.  Note that $(\tilde{ \varphi} t_R)$ transforms as a $(2,\frac16 )$ under the electroweak interactions, like the SM left-handed top-bottom doublet; the operator (\ref{eq:smtl-op}) containing this combination of fields arises when the vector fermion $\Psi$ state being integrated out carries those SM-like quantum numbers. Models in the literature that contain such operators include \cite{He:1999vp,Pomarol:2008bh,Chivukula:2006cg,Belyaev:2009ve,Chivukula:2009ck,Chivukula:2011ag}. Likewise, $( \varphi t_R)$ transforms like an exotic $(2, \frac76 )$ and the associated operator (\ref{eq:extl-op}) results when the new vector fermion has those exotic quantum numbers; related models include \cite{Pomarol:2008bh,SekharChivukula:2009if}.  Both of these operators will affect $\alpha T$ at one loop because they alter the $Z$ but not the $W$ propagator (since the $W$ does not couple to $t_R$).  Neither affects $R_b$ at tree level since they alter the $Z \bar{t}_R t_R$ vertex rather than the $Z \bar{b}b$ vertex; however they could affect $R_b$ at one loop through diagrams with an internal top quark coupled to the decaying $Z$ boson.   Both of these operators will be worth exploring further.

When a right-handed state mixing with $t_R$ has been integrated out, as in \cite{Pomarol:2008bh,Dobrescu:1997nm,Chivukula:1998wd}, one obtains
\begin{equation}
{\cal O}_3 \equiv (\bar{q}_L\tilde{ \varphi}) i \slashed{D} (\tilde{ \varphi}^\dagger q_L) \label{eq:smtr-op}
\end{equation}
Here, $(\tilde{ \varphi}^\dagger q_{L})$ has the quantum numbers of the SM $t_R$ quark.  This operator will affect both the $Z \bar{t}_L t_L$ and $W \bar{t}_L b_L$ couplings (and may therefore impact single-top production).  Because it does not also affect the $Z b_L b_L$ coupling, it {\it will} alter $\alpha T$ at one loop -- but {\it will not} affect $R_b$ at tree level; it should generally impact $R_b$ at the one-loop level through diagrams with an internal top quark.  This operator also merits further study.

In contrast, if a right-handed state mixing with $b_R$ is integrated out, the resulting operator 
\begin{equation*}
{\cal O}_4 \equiv (\bar{q}_L{ \varphi}) i \slashed{D} ({ \varphi}^\dagger q_L) 
\end{equation*}
includes  $(\varphi^\dagger q_L)$ which has the quantum numbers of the SM $b_R$.  This operator affects the $Z \bar{b}_L b_L$ vertex at tree level and, as such, is very tightly constrained.  Therefore, this operator does not warrant further study at present.

Finally, if custodial and flavor symmetry protect the interactions of the SM-like top-bottom doublet, then integrating out the states mixing with $t_R$ and $b_R$, as in \cite{Chivukula:2006cg,Belyaev:2009ve,Chivukula:2009ck,Chivukula:2011ag}, yields the custodially-symmetric operator
\begin{equation*}
{\cal O}_5 \equiv (\bar{q}_L \Phi) i \slashed{D} (\Phi^\dagger q_L)
\end{equation*}
in which $\Phi \equiv (\tilde{\varphi}, \varphi)$ and $\tilde{\varphi} = i \sigma_2 \varphi^*$.
This operator does not contribute to $\alpha T$ at all.  In the presence of a flavor symmetry, the operator will not alter $R_b$ either \cite{Chivukula:2006cg}.   If one, instead, integrates out heavy vector states mixing with $q_L$, the resulting operator
\begin{equation*}
{\cal O}_6 \equiv (\bar{q}_R \lambda^\dagger \Phi^\dagger)  i D\!\!\!\!/\, (\Phi \lambda q_R)
\end{equation*}
encodes isospin breaking in the Yukawa coupling matrix $\lambda \equiv Diag(\lambda_t, \lambda_b)$.  This will affect the $Z \bar{t}_R t_R$ and $Z \bar{b}_R b_R$ couplings -- with the latter effect suppressed by $\lambda_b^2 / \lambda_t^2$ to a level that is unilkely to be interesting.  A $W \bar{t}_R b_R$ coupling can be induced, as occurs in the 3-site model, but this is again suppressed by $m_b / m_t$ and is unlikely to yield interesting limits.  We will not explore either of these operators further.

Our discussion has identified three operators of potential phenomenological interest $({\cal O}_1, {\cal O}_2, {\cal O}_3)$, and shown that they arise from mixing between fundamental top and bottom quarks with particular kinds of (potentially composite) vector fermions.  In the next sections of the paper, we will introduce specific models that explore the ideas raised by our effective field theory analysis.   Section~\ref{sec:explicit} proposes a pair of illustrative models in which the Yukawa interactions explicitly break the $SU(2)_c$ custodial symmetry of the symmetry breaking sector. In the first model the right-handed top is mixed with a heavy vector singlet, so that this model explores operator (\ref{eq:smtr-op}), whereas in the second model the left-handed top-bottom doublet is mixed with a heavy vector doublet so that one instead explores operator (\ref{eq:smtl-op}). Each of these models include all terms consistent with the gauge symmetries, and each is a renormalizable four-dimensional gauge theory.  We will find that neither kind of mixing leads, on its own, to a more viable phenomenology -- and this implies that  models introducing both kinds of mixing at once will be similarly unsuccessful.

In section~\ref{sec:soft} we introduce models incorporating a full $O(4)$ symmetry in the Yukawa sector; again, each model includes all terms consistent with the gauge symmetries.\footnote{These last two models are not truly renormalizable: the $SU(2) \times
U(1)$ electroweak gauge interactions do not respect the $SU(2)_L \times SU(2)_R \times P_{LR}$ global symmetries assumed
for the Yukawa couplings. However, we find that these effects are negligible so long as the underlying strong dynamics respects $SU(2)_L \times SU(2)_R \times P_{LR}$.}   We first consider a model in which the left-handed top-bottom doublet is mixed with a heavy vector bi-doublet of $SU(2)_L\times SU(2)_R$; integrating out the heavy vectors would give rise to the third interesting operator identified in our effective field theory analysis (\ref{eq:extl-op}).  We find that this particular model has limitations, but also suggests a possible solution.  We then instantiate that solution in a particularly economical way by  embedding the top-bottom left-handed doublet itself in an $SU(2)_L\times SU(2)_R$ bi-doublet \cite{SekharChivukula:2009if}, $(q_L,\Psi_L)$, while mixing the right-handed top with a heavy vector singlet. For each of the models considered in this section,  in the limit of zero gauge couplings, the $O(4) $ symmetry is softly broken due to a Dirac mass term for the vector fermions. Moreover, these two models contain $O(3)_c$ triplet fermions including states of charge $+2/3$
and $T_{3L}=-1/2$ which, after electroweak symmetry breaking, mix with the top-quark. We find that only models with these custodial triplet fermions feature both positive and negative contributions to $\hat{T}$ and $g_{Lb}$, leading to regions of the parameter space in which agreement with experiment is at the $1\sigma$ level.

\section{Models Producing Operators ${\cal O }_1$ and ${\cal O}_3$:  Breaking Custodial Symmetry via Yukawa Couplings}\label{sec:explicit}

In this section we examine a pair of scenarios in which the custodial symmetry is explicitly broken by Yukawa couplings.  We will find that neither of these directions in model building, on their own, provides improved agreement with data on $\hat{T}$ and $g_{Lb}$, compared with the SM.  This is because in both models the corrections to both parameters turn out to be uniformly positive and correlated with one another.

\subsection{Weak Singlet Compositeness}\label{sec:2A}

A vector-like custodial/weak singlet fermion $t_1$ has the same charges as the right-handed top of the SM.  Adding  such a field to the top-sector Lagrangian gives\footnote{Here and in the following we neglect generation mixings.}
\begin{equation}
{\cal L}_{\rm top} = 
\bar{q}_{0L} i \slashed{D} q_{0L} 
+ \bar{t}_{0R} i \slashed{D} t_{0R}
+ \bar{t}_1 i \slashed{D} t_1
-M_t \bar{t}_1 t_1
-\mu_t \left(\bar{t}_{0R} t_{1L} + {\rm h.c.}\right)
-y_t\left(\bar{q}_{0L} \widetilde{\varphi}\ t_{1R} + {\rm h.c.}\right) \ ,
\label{eq:Lsinglet}
\end{equation}
where $q_{0L}\equiv(t_{0L}, b_{0L})$ is the elementary left-handed top-bottom doublet, $\varphi$ is the $Y=1/2$ Higgs doublet, and $\widetilde{\varphi}\equiv i\sigma^2\varphi^\ast$ is the $Y=-1/2$ version of $\varphi$.  While a Yukawa term mixing $q_{0L}$ with $t_{0R}$ is also possible, it can always be removed by redefining $t_{0R}$ and $t_{1R}$. For our analysis the most relevant feature of this Lagrangian is the (explicit) hard breaking of custodial symmetry in the Yukawa sector (i.e. the fact that $y_t \neq y_b$).  In fact, because the right-handed bottom quark does not contribute significantly to any isospin-violating processes, we will simply set $y_b = 0$ and ignore the right-handed bottom quark altogether.  Note that in the effective theory resulting from taking the limit $M_t \to \infty$ in Eq. (\ref{eq:Lsinglet}) we generate an operator of the form $(\bar{q}_{0L}\tilde{\varphi}) i\slashed{D} (\tilde{\varphi}^\dagger q_{0L})$, that is, an operator of the form of ${\cal O}_3$ in Eq.~(\ref{eq:smtr-op}). In other words, below the mass of the partner fermion state, the effective theory includes both SM physics and additional effects from ${\cal O}_3$.

\begin{figure}[!t]
\includegraphics[bb=5.0in 6.0in 3.0in 11.0in]{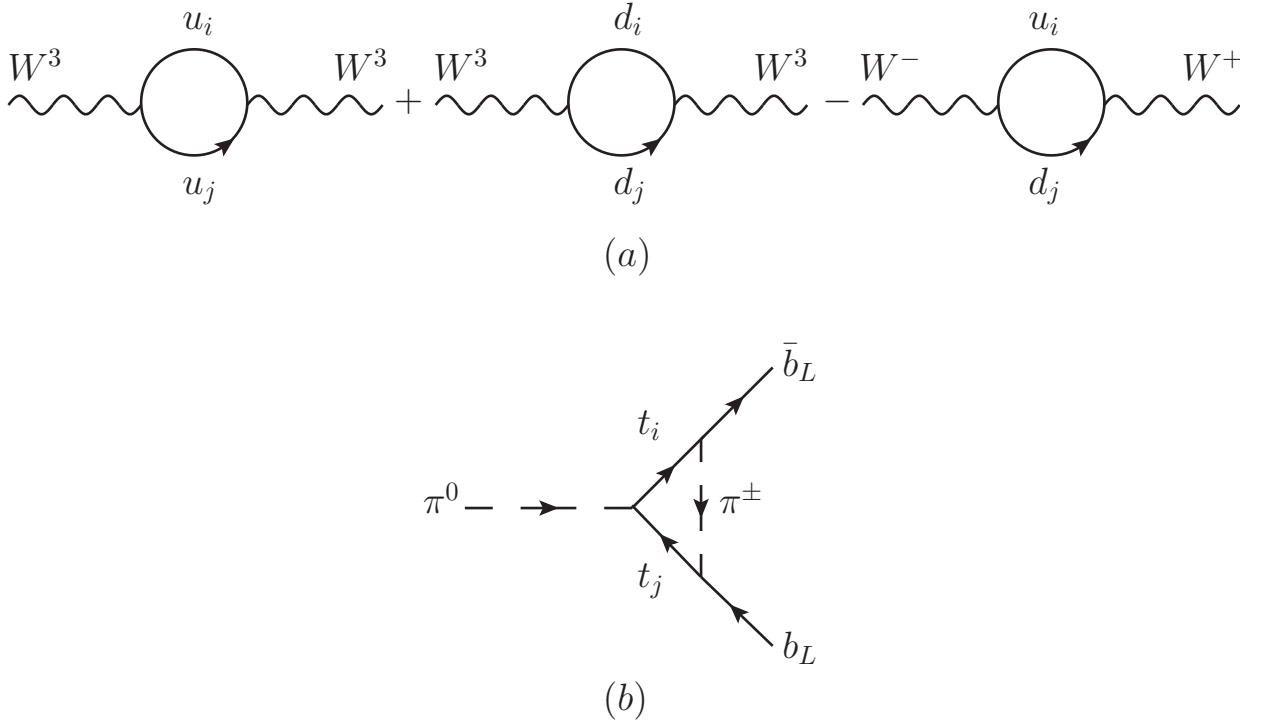}
\caption{(a) Diagrams contributing to the $\hat{T}$ parameter. Here $u_i$ and $d_i$ are up-type and down-type fermions, respectively, in the top sector (the lighter fermions give negligible contributions). (b) Diagram contributing to the left-handed $Z b \bar{b}$ coupling in gaugeless limit: $\pi^0$ and $\pi^\pm$ are the Goldstone bosons eaten by $Z$ and $W$, respectively.}
\label{fig:diagrams}
\end{figure}

The mass terms in Eq.~(\ref{eq:Lsinglet}) may be written in matrix notation as
\begin{equation}
{\cal L}_{\rm mass} = -
\left(\begin{array}{cc} t_{0L} & t_{1L} \end{array}\right)
\left(\begin{array}{cc} 0 & \hat{m}_t \\ \mu_t & M_t \end{array}\right)
\left(\begin{array}{c} t_{0R} \\ t_{1R} \end{array}\right)
+ {\rm h.c.} \ ,
\end{equation}
where
\begin{equation}
\hat{m}_t \equiv \displaystyle{\frac{y_t v}{\sqrt{2}}} \ .
\label{eq:mthat}
\end{equation}
Diagonalization shows how to rewrite the gauge eigenstates in terms of the mass eigenstates:
\begin{equation}
\left(\begin{array}{c}t_{0L} \\ t_{1L}\end{array}\right)=
\left(\begin{array}{cc}\cos\theta_L & \sin\theta_L \\ -\sin\theta_L & \cos\theta_L \end{array}\right)
\left(\begin{array}{c}t_L \\ T_L\end{array}\right) \ ,\quad\quad
\left(\begin{array}{c}t_{0R} \\ t_{1R}\end{array}\right)=
\left(\begin{array}{cc}\cos\theta_R & -\sin\theta_R \\ \sin\theta_R & \cos\theta_R \end{array}\right)
\left(\begin{array}{c}t_R \\ T_R\end{array}\right) \ ,
\label{eq:diagtop}
\end{equation}
where $t$ and $T$ are, respectively, the ordinary top and its heavy partner,  and the mixing factors are:
\begin{eqnarray}
\sin\theta_L =\frac{\cos^2\beta}{\sin\beta}\ \frac{m_t}{M_t}
\left[1-\displaystyle{\frac{2\cos^2\beta\ m_t^2}{M_t^2}}+\displaystyle{\frac{\cos^4\beta\ m_t^4}{\sin^2\beta\ M_t^4}}\right]^{-1/2} \ , \quad
\sin\theta_R = \left[1+\left(\tan\beta-\frac{\cot\beta\ m_t^2}{M_t^2}\right)^{-2}\right]^{-1/2} \ .
\end{eqnarray}
These rotation angles are conveniently expressed in terms of $\beta$, which measures the amount of mixing of $t_{0R}$ with the vector fermion $t_1$:
\begin{equation}
\tan\beta\equiv \displaystyle{\frac{\mu_t}{M_t}}\ .
\end{equation}
We have also eliminated the parameter $\hat{m}_t$ in favor of the physical top-quark mass, $m_t$:
\begin{equation}
\hat{m}_t = \frac{m_t}{\sin\beta}\sqrt{\frac{M_t^2-\cos^2\beta\ m_t^2}{M_t^2-\cot^2\beta\ m_t^2}} \ .
\end{equation}
In terms of $\beta$, $M_t$, and $m_t$, the mass of the heavy $T$ state is:
\begin{equation}
M_T = \frac{M_t}{\cos\beta}\sqrt{\frac{M_t^2-\cos^2\beta\ m_t^2}{M_t^2-\cot^2\beta\ m_t^2}} \ .
\end{equation}
The SM limit of this theory  is obtained by sending $\mu_t$ to infinity, independent of $M_t$, and hence sending $\sin\beta\to 1$; in this limit, the left-handed top no longer mixes with new states so the light top eigenstate behaves like the SM top.

\begin{figure}[!t]
\includegraphics[width=7cm,height=5cm]{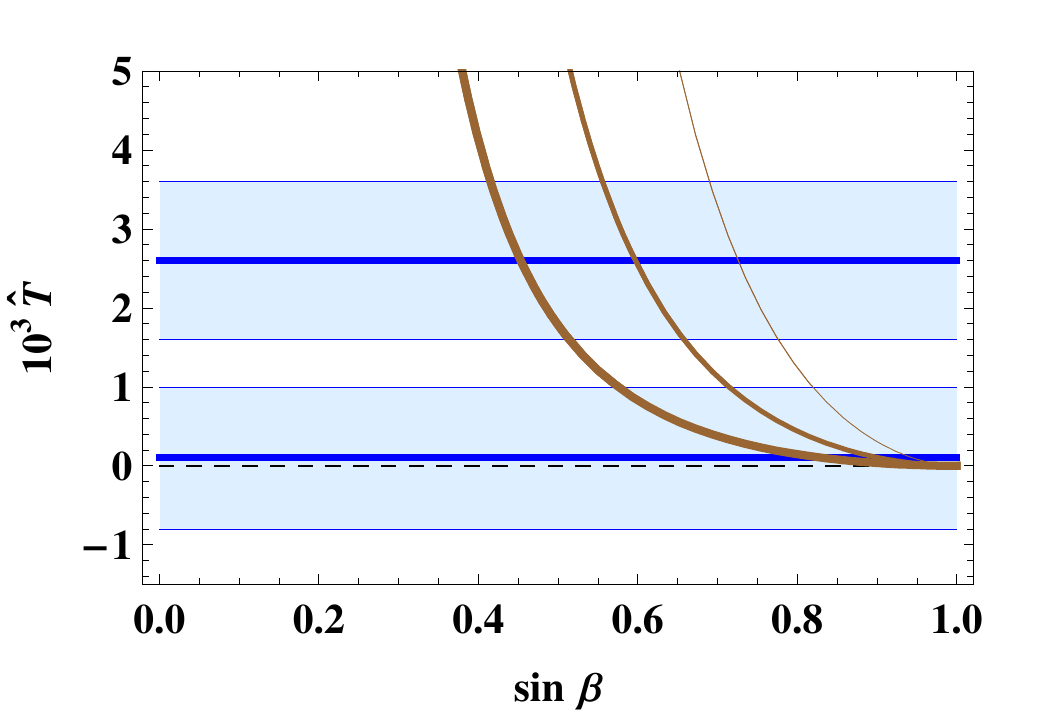}
\includegraphics[width=7cm,height=5cm]{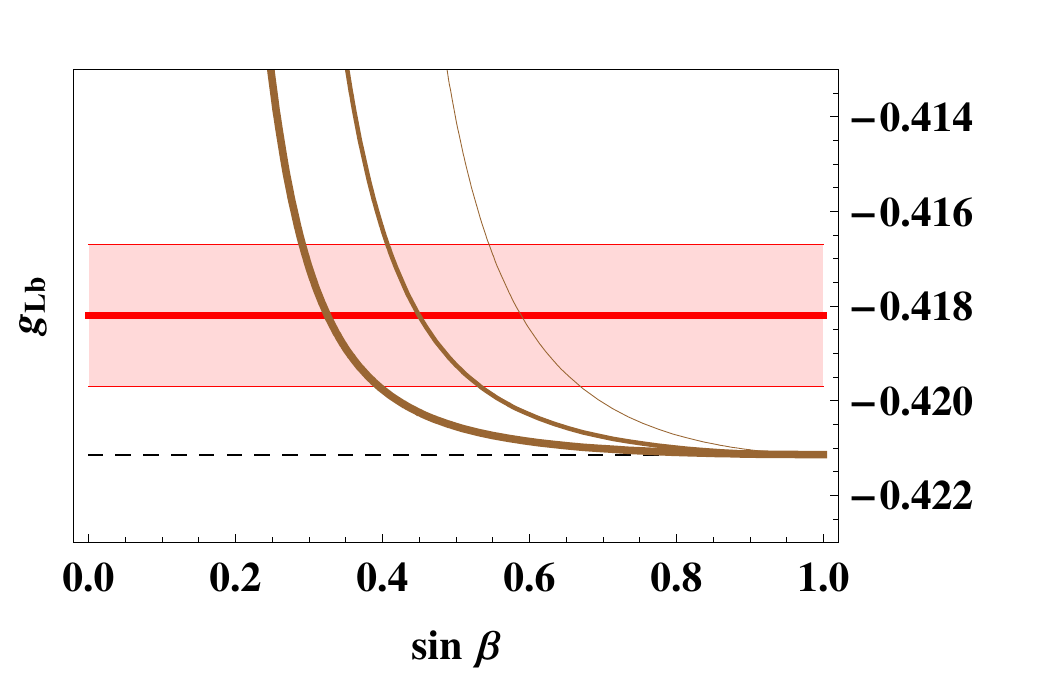}
\caption{$\hat{T}$ (left) and $\delta g_{Lb}$ (right) in the model of Eq.~(\ref{eq:Lsinglet}). The thin to thick curves are for $M_t=$ 0.5, 1, and 2 TeV, respectively, whereas the dashed lines are the SM predictions. In the plot for $\hat{T}$ the lower (upper) experimental $1\sigma$ band is for $m_H=115$ GeV ($m_H=800$ GeV) \protect\cite{Barbieri:2004qk}. The experimental $1\sigma$ band for $\delta g_{Lb}$ has a negligible dependence on the Higgs mass \protect\cite{:2005ema}. }
\label{fig:plotsinglet}
\end{figure}

With the fermion fields diagonalized it is straightforward to compute the dominant contributions to $\hat{T}$ and $\delta g_{Lb}$. The corrections to $\hat{T}$ arise from the diagrams of Fig.~\ref{fig:diagrams}(a), where $u_i$ and $d_i$ are up-type and down-type fermions, respectively, in the top sector. In the present case $u_i=t,T$ and $d_i=b$. The dominant corrections to $g_{Lb}$ can be computed in the gaugeless limit \cite{Lytel:1980zh,Barbieri:1992nz,Barbieri:1992dq,Oliver:2002up}, and are given by the diagram of Fig.~\ref{fig:diagrams}(b) where $t_i$ and $t_j$ are $t,T$. The values of $\hat{T}$ and $\delta g_{Lb}$ (defined\footnote{The $Z$ coupling to left-handed $b$ quarks is written as $\dfrac{e}{\sin\theta_W \cos\theta_W} \left(-\dfrac12 + \delta g_{Lb} + \dfrac13 \sin^2\theta_W\right)$. } as the new-physics contribution to $g_{Lb}$) turn out to be positive and correlated; we find, in agreement with \cite{Carena:2004zn}, the expressions:
\begin{equation}
\frac{\hat{T}}{\hat{T}^{\rm SM}}=\frac{\delta g_{Lb}}{\delta g_{Lb}^{\rm SM}}
=\left(4\log\frac{M_t/\cos\beta}{m_t}+\frac{1}{\sin^2\beta}-3\right)
\frac{\cos^4\beta\ m_t^2}{\sin^2\beta\ M_t^2}\left[1+{\cal O}\left(\frac{m_t^4}{M_t^4}\right)\right] \ ,
\label{eq:singlet}
\end{equation}
where
\begin{eqnarray}
\hat{T}^{\rm SM} = \frac{3 m_t^2}{16\pi^2 v^2} \ , \quad
\delta g_{Lb}^{\rm SM} = \frac{m_t^2}{16\pi^2 v^2} \ .
\end{eqnarray}
The contribution to the $\hat{S}$ parameter is also positive (for $M_t/\cos\beta\gtrsim 600$ GeV), but numerically much smaller than $\hat{T}$:
\begin{equation}
\hat{S}=\frac{g^2}{96\pi^2}\left(4\log\frac{M_t/\cos\beta}{m_t}-5\right)
\frac{\cos^4\beta\ m_t^2}{\sin^2\beta\ M_t^2}\left[1+{\cal O}\left(\frac{m_t^4}{M_t^4}\right)\right] \ .
\label{eq:Shatsinglet}
\end{equation}

In the left-hand pane of  Fig.~\ref{fig:plotsinglet} we plot $\hat{T}$ as a function of $\sin\beta$ for $M_t=$ 0.5, 1, and 2 TeV (thin to thick curves, respectively), together with the SM predictions (dashed lines). The experimental value of $\hat{T}$ depends on the Higgs mass: the lower (upper) $1\sigma$ band is for $m_H=115$ GeV ($m_H=800$ GeV). These bounds are taken for an arbitrary value of $\hat{S}$: the latter is much smaller than $\hat{T}$, and within $1\sigma$ for both light and heavy Higgs. We see that new-physics contributions to $\hat{T}$ in this model reduce agreement with the light-Higgs limit of the SM and favor a heavier Higgs.

The right-hand pane of  Fig.~\ref{fig:plotsinglet} shows $\delta g_{Lb}$ as a function of $\sin\beta$ for the same $M_t$ as before;  here, the experimental value is nearly independent of $m_H$.  The positive correction to $\delta{g_{Lb}}$ in Eq.~(\ref{eq:singlet}) does tend to push $g_{Lb}$ in the direction required for better agreement with experiment. However, as alluded to above, the values of $\sin\beta$ for which $\delta{g_{Lb}}$ is within $1\sigma$ of the data disagree with $\hat{T}$ for a light Higgs by more than $2\sigma$.  And while those values of $\sin\beta$ do yield agreement at the $2\sigma$ level for $\hat{T}$ if a heavy Higgs is assumed, overall the model does not surpass the SM in its agreement with the broader electroweak precision data.

The chief difficulty lies with $\delta g_{Rb}$.  In this kind of model, the large $m_t$ - $m_b$ splitting arises from the fact that $y_t >> y_b$; and we have worked in the limit where $y_b$ is simply set to zero.  So in general, this kind of model would predict $\delta g_{bR} << \delta g_{bL}$ and in the limit we adopt, $\delta g_{bR} = 0$.  However, as mentioned earlier, the measured left-handed and right-handed $Z b \bar{b}$ couplings have a strong and positive correlation, and the SM predictions are both approximately $2\sigma$ below their expectation values. Therefore, a positive correction to $g_{Lb}$ without a corresponding positive correction to $g_{Rb}$ cannot restore agreement with experiment.

\subsection{Weak Doublet Compositeness}
\label{sec:2B}

A vector-like custodial/weak doublet, $q_1\equiv(t_1,b_1)$, has the same charges as the SM left-handed top-bottom doublet, here denoted by $q_{0L}\equiv(t_{0L},b_{0L})$. Including $q_1$ in the SM top-sector yields the Lagrangian
\begin{equation}
{\cal L}_{\rm top} = 
\bar{q}_{0L} i \slashed{D} q_{0L} 
+ \bar{t}_{0R} i \slashed{D} t_{0R}
+ \bar{q}_1 i \slashed{D} q_1
-M_q \bar{q}_1 q_1
-\mu_q \left(\bar{q}_{0L} q_{1R} + {\rm h.c.}\right)
-y_t\left(\bar{q}_{1L} \widetilde{\varphi}\ t_{0R} + {\rm h.c.}\right) \ .
\label{eq:Ldoublet}
\end{equation}
A Yukawa term mixing $q_{0L}$ and $t_{0R}$ is also possible, but can always be removed by redefining $q_{0L}$ and $q_{1L}$. As in the composite singlet model considered above, in this model the breaking of custodial $O(4) $ in the Yukawa sector is explicit.  This time, in the effective theory resulting from taking the limit $M_q \to \infty$ in Eq. (\ref{eq:Ldoublet}) we generate an operator of the form $(\bar{t}_{0R}\tilde{\varphi}^\dagger) i\slashed{D} (\tilde{\varphi} t_{0R})$, that is, an operator of the form of  ${\cal O}_1$ in Eq.~(\ref{eq:smtl-op}); below the mass of the partner fermion state, the effective theory includes both SM physics and additional effects from ${\cal O}_1$.

\begin{figure}[!t]
\includegraphics[width=7cm,height=5cm]{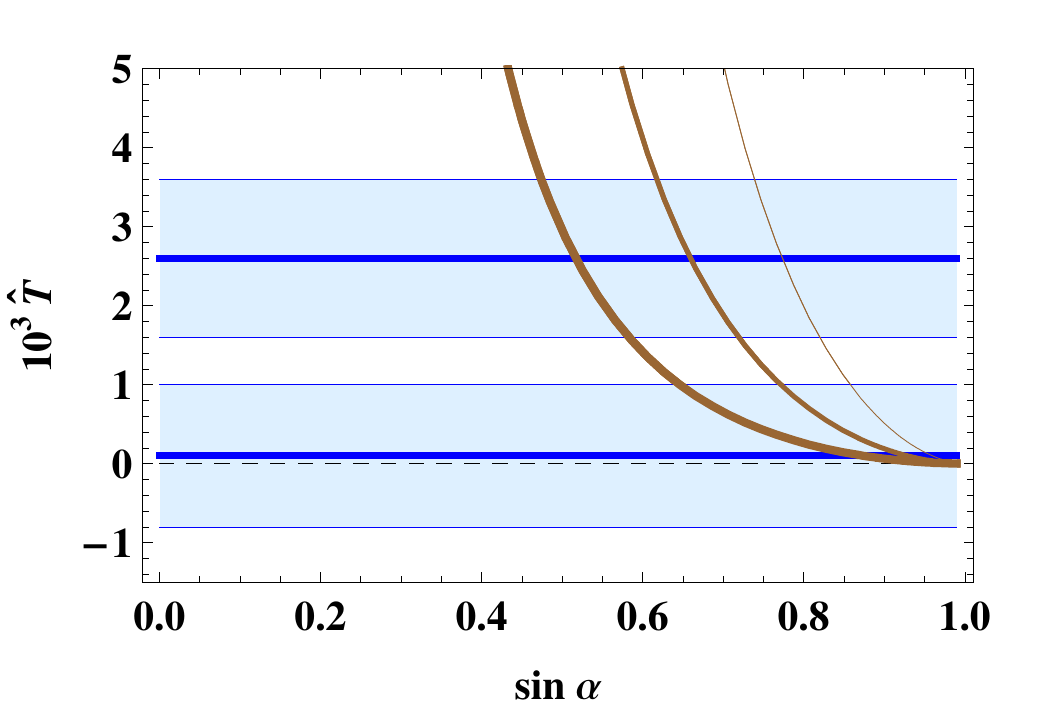}
\includegraphics[width=7cm,height=5cm]{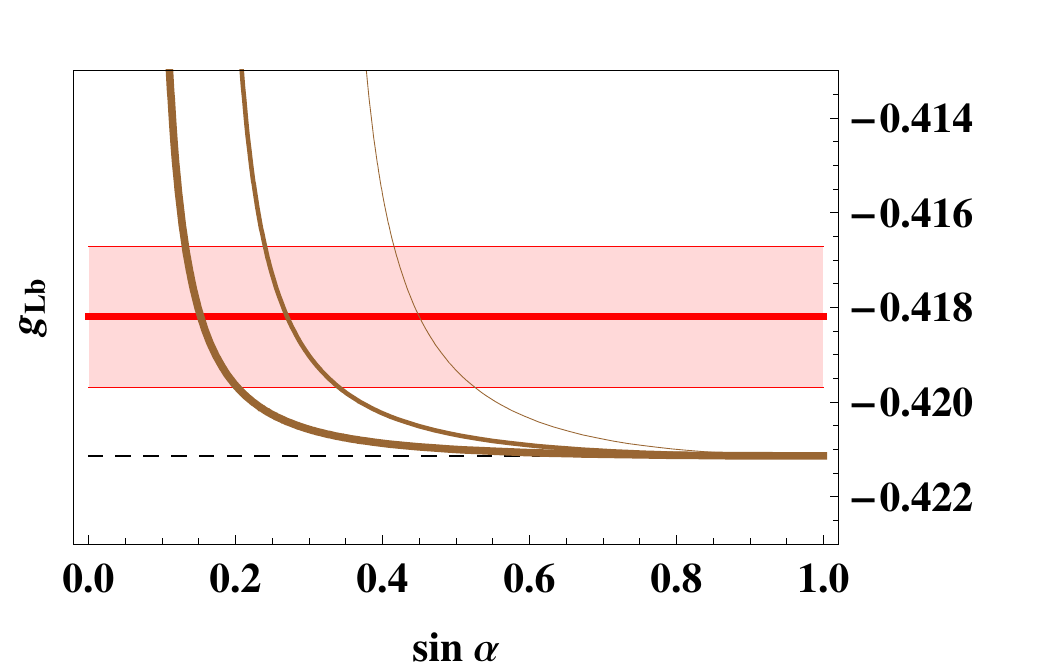}
\caption{$\hat{T}$ (left) and $\delta g_{Lb}$ (right) in the model of Eq.~(\ref{eq:Ldoublet}). The thin to thick curves are for $M_q=$ 0.5, 1, and 2 TeV, respectively, whereas the dashed lines are the SM predictions. In the plot for $\hat{T}$ the lower (upper) experimental $1\sigma$ band is for $m_H=115$ GeV ($m_H=800$ GeV) \protect\cite{Barbieri:2004qk}. The experimental $1\sigma$ band for $\delta g_{Lb}$ has a negligible dependence on the Higgs mass. \cite{:2005ema}}
\label{fig:plotdoublet}
\end{figure}

The mass Lagrangian is
\begin{equation}
{\cal L}_{\rm mass} = -
\left(\begin{array}{cc} t_{0L} & t_{1L} \end{array}\right)
\left(\begin{array}{cc} 0 & \mu_q \\ \hat{m}_t & M_q \end{array}\right)
\left(\begin{array}{c} t_{0R} \\ t_{1R} \end{array}\right)
-\left(\mu_q b_{0L} + M_q b_{1L}\right)b_{1R} + {\rm h.c.} \ ,
\end{equation}
where $\hat{m}_t \equiv y_q v / \sqrt{2}$. Diagonalization of the top fields is still expressed by Eq.~(\ref{eq:diagtop}), where now
\begin{equation}
\sin\theta_L = \frac{\tan\alpha-{\displaystyle\frac{\cot\alpha\ m_t^2}{M_q^2}}}
{\sqrt{1+\left(\tan\alpha-{\displaystyle\frac{\cot\alpha\ m_t^2}{M_q^2}}\right)^2}} \ ,\quad\quad
\cos\theta_R = \frac{m_t/M_q}{\sqrt{{\displaystyle \frac{\tan^2\alpha}{\cos^2\alpha}-\frac{2\tan^2\alpha\ m_t^2}{M_q^2}+\frac{m_t^4}{M_q^4}}}} \ .
\label{eq:angles}
\end{equation}
Here $\alpha$ measures the amount of mixing of $q_{0L}$ with the new vector fermion $q_1$,
\begin{equation}
\tan\alpha\equiv \displaystyle{\frac{\mu_q}{M_q}}\ .
\label{eq:alpha}
\end{equation}
In deriving Eq.~(\ref{eq:angles}) we have rewritten $\hat{m}_t$ in terms of the top-quark mass $m_t$,
\begin{equation}
\hat{m}_t = \frac{m_t}{\sin\alpha}\sqrt{\frac{M_q^2-\cos^2\alpha\ m_t^2}{M_q^2-\cot^2\alpha\ m_t^2}} \ ,
\end{equation}
and in terms of those same variables, the heavy top mass is
\begin{equation}
M_T = \frac{M_q}{\cos\alpha}\sqrt{\frac{M_q^2-\cos^2\alpha\ m_t^2}{M_q^2-\cot^2\alpha\ m_t^2}} \ .
\end{equation}
In this model, because the new vector multiplet includes a partner for the b-quark, $b_{0L}$ (the gauge eigenstate with SM quantum numbers) mixes with the new vector fermion $b_{1L}$. The mass eigenstates are diagonalized by
\begin{equation}
\left(\begin{array}{c}b_{0L} \\ b_{1L}\end{array}\right)=
\left(\begin{array}{cc}\cos\alpha & \sin\alpha \\ -\sin\alpha & \cos\alpha \end{array}\right)
\left(\begin{array}{c}b_L \\ B_L\end{array}\right) \ ,
\label{eq:bottom}
\end{equation}
where the mass of the heavy $B$ quark is
\begin{equation}
m_B = \frac{M_q}{\cos\alpha} \ .
\label{eq:Bmass}
\end{equation}
In this case, the SM limit corresponds to $\mu_q\to\infty$ with finite $M_q$ (regardless of the particular value of $M_q$), which implies $\sin\alpha\to 1$;  in this limit, the left-handed top no longer mixes with new states and the light eigenstate behaves like the SM top.

The one-loop contributions to $\hat{T}$ and $\delta g_{Lb}$ are still given by the diagrams of Fig.~\ref{fig:diagrams}(a) and \ref{fig:diagrams}(b), respectively, with $u_i, t_i=t,T$, and $d_i=b,B$. This yields
\begin{eqnarray}
&&\hat{T} = \hat{T}^{\rm SM}\left(8\log\frac{M_q/\cos\alpha}{m_t}+\frac{4}{3\sin^2\alpha}-\frac{22}{3}\right)
\frac{\cos^4\alpha\ m_t^2}{\sin^2\alpha\ M_q^2}\left[1+{\cal O}\left(\frac{m_t^4}{M_q^4}\right)\right] \ , \\
&&\delta g_{Lb} = \delta g_{Lb}^{\rm SM}\ \log\frac{M_q/\cos\alpha}{m_t}
\frac{\cos^4\alpha\ m_t^2}{\sin^2\alpha\ M_q^2}\left[1+{\cal O}\left(\frac{m_t^4}{M_q^4}\right)\right] \ .
\end{eqnarray}
The $\hat{S}$ parameter is, again, numerically much smaller than $\hat{T}$:
\begin{equation}
\hat{S}=\frac{g^2}{96\pi^2}\left(8\log\frac{M_q/\cos\alpha}{m_t}-7\right)\frac{\cos^4\alpha\ m_t^2}{\sin^2\alpha\ M_q^2}
\left[1+{\cal O}\left(\frac{m_t^4}{M_q^4}\right)\right] \ .
\end{equation}

In Fig.~\ref{fig:plotdoublet} we plot $\hat{T}$ and $\delta g_{Lb}$, using the same notation as Fig.~\ref{fig:plotsinglet}. The disagreement with data is even worse than in the composite singlet case because $\delta g_{Lb}$ is still positive but grows more slowly with decreasing $\sin\alpha$ than $\hat{T}$ does.  Values of $\sin\alpha$ for which $\delta g_{Lb}$ is within $1\sigma$ of the data give values of $\hat{T}$ that are several standard deviations larger than the experimental value for any $m_H$.  Moreover, the problems with $\delta g_{Rb}$ persist; since $y_b = 0$ the model predicts no shift in $g_{Rb}$ from the SM value.

\section{Models Producing Operator ${\cal O}_2$: Soft Breaking of Custodial Symmetry}\label{sec:soft}

In this section, we explore models incorporating a softly-broken extended custodial symmetry.  We show that adding a vector fermion bi-doublet to the SM-like fields introduces new contributions to $g_{Lb}$ that largely cancel one another, leaving $g_{Lb}$ close to the SM prediction. The bi-doublet contribution to $\hat{T}$ is negative: as a consequence, both $g_{Lb}$ and $\hat{T}$ lie below the experimental values.  This is a different pattern than we found in the models of section \ref{sec:explicit}, but is not satisfactory on its own.

It is then natural to expect that including an additional vector-like singlet in an $O(4) $ symmetric fashion can help to achieve agreement with experiment, since the singlet contributions to $g_{Lb}$ and $\hat{T}$ are both positive, as found in section~\ref{sec:2A}. We will show that this is indeed the case, and that there are regions of the parameter space in which the agreement with experiment is at the $1\sigma$ level, for either a light or a heavy Higgs.

\subsection{Heavy Vector Bi-Doublet}\label{sec:3A}

Let $Q_1$ be a heavy vector-like bi-doublet of $SU(2)_L\times SU(2)_R$,
\begin{eqnarray}
Q_1\equiv\left(\begin{array}{cc} q_1 & \Psi_1 \end{array}\right)=\left(\begin{array}{cc}t^q_1 & \Omega_1 \\ b_1 & t^\Psi_1\end{array}\right) \ .
\end{eqnarray}
Note that the field $Q_1$ includes both $O(3)_c$ singlet and triplet components, and that the field $\Omega$ has electric charge $+5/3$ and $T_{3L} = +1/2$ while the field $t^\Psi_1$ has electric charge $+2/3$ and $T_{3L}=-1/2$  \cite{SekharChivukula:2009if}.  In addition, collect the Higgs doublet $\varphi$ together with its hypercharge conjugate $\widetilde{\varphi}$ in another $SU(2)_L\times SU(2)_R$ bi-doublet:
\begin{eqnarray}
\Phi =\left(\widetilde{\varphi},\varphi\right) \ .
\end{eqnarray}

Consider the following top-sector Lagrangian that includes the new vector fermion $Q_1$, and the Higgs bi-doublet $\Phi$ along with the SM-like elementary doublet, $q_{0L}\equiv (t_{0L},b_{0L})$ and the SM-like elementary singlet $t_{0R}$:
\begin{equation}
{\cal L}_{\rm top} = \bar{q}_{0L} i \slashed{D} q_{0L} 
+ \bar{t}_{0R} i \slashed{D} t_{0R}
- {\rm Tr}\ \bar{Q}_1 i \slashed{D} Q_1
- \mu_q \left(\bar{q}_{0L} q_{1R} + {\rm h.c.}\right)
- M_q\ {\rm Tr}\ \bar{Q}_1 Q_1
- y_t {\rm Tr}\left(\bar{Q}_{1L}\Phi\ t_{0R} + {\rm h.c.} \right) \ .
\label{eq:Lbidoublet1}
\end{equation}
In the effective theory resulting from taking the limit $M_q \to \infty$ in Eq. (\ref{eq:Lsinglet}) we generate an operator of the form $(\bar{t}_{0R}{\varphi}) i\slashed{D} ({\varphi}^\dagger t_{0R})$, that is, an operator of the form of ${\cal O}_2$ in Eq.~(\ref{eq:extl-op}).
In the limit of zero gauge couplings, and for $\mu_q=0$, this Lagrangian features an $SU(2)_L\times SU(2)_R\times P_{LR}\sim O(4)$ global symmetry, which is spontaneously broken to custodial $SU(2)_c\times P_{LR}\sim O(3)_c$. This symmetry is reflected in the identical $y_t$ coefficients for $\bar{q}_{1L}\widetilde{\varphi}\ t_{0R}$ and $\bar{\Psi}_{1L}\varphi\ t_{0R}$. The $P_{LR}$ transformations are
\begin{eqnarray}
Q_1 \to - \epsilon\ Q_1^T\ \epsilon\ , \quad  \Phi \to - \epsilon\ \Phi^T\ \epsilon\ , \quad q_{0L}\to q_{0L}\ , \quad t_{0R}\to t_{0R} \ ,
\end{eqnarray}
where
\begin{eqnarray}
\epsilon=\left(\begin{array}{cc} 0 & 1 \\ -1 & 0\end{array}\right) \ .
\end{eqnarray}

Switching on $\mu_q$ introduces a source of soft $O(4) $ breaking that does not generate $O(4) $-breaking counterterms. In contrast, switching on the gauge interactions leads to a hard breaking of custodial $O(4)$ (for instance, the weak interaction does not respect $P_{LR}$) and this generates $O(4) $-breaking counterterms. We estimate the impact of this breaking pattern on the quantities of interest to us as follows. If the Yukawa interactions are like those of Eq.~(\ref{eq:Lbidoublet1}) at some higher-energy ``compositeness scale'' $\Lambda$, then at the electroweak scale the coefficients of $\bar{q}_{1L}\widetilde{\varphi}\ t_{0R}$ and $\bar{\Psi}_{1L}\varphi\ t_{0R}$ will differ by a quantity $\Delta y_t$ which can be estimated to be of order
\begin{equation}
\Delta y_t \sim \frac{y_t\ g^2}{16\pi^2}\log\frac{\Lambda^2}{v^2} \nonumber \ .
\end{equation}
For example, taking $\Lambda=$10 TeV gives a very small correction, $\Delta y_t/y_t\sim 0.02$. Therefore, we can ignore the running of $\Delta y_t$, and take custodial $O(4) $ to be exact in the Yukawa sector at the electroweak scale\footnote{The Yukawa interaction $\bar{q}_{0L}\widetilde{\varphi}\ t_{0R}$ is not generated at loop level, because of a $Z_2$ symmetry under which $Q_1\to -Q_1$ and $\Phi\to -\Phi$. The latter is only broken softly by the $\mu_q$ term.}.

The mass Lagrangian may be summarized in matrix form as
\begin{equation}
{\cal L}_{\rm mass} = -
\left(\begin{array}{ccc} t_{0L} & t^q_{1L} & t^\Psi_{1L} \end{array}\right)
\left(\begin{array}{ccc} 0 & \mu_q & 0 \\ \hat{m}_t & M_q & 0 \\ \hat{m}_t & 0 & M_q \end{array}\right)
\left(\begin{array}{c} t_{0R} \\ t^q_{1R} \\ t^\Psi_{1R} \end{array}\right)
-\left(\mu_q b_{0L} + M_q b_{1L}\right)b_{1R} - M_q\bar{\Omega}_{1L}\Omega_{1R}+ {\rm h.c.} \ ,
\end{equation}
where $\hat{m}_t$ is still given by Eq.~(\ref{eq:mthat}). As usual, we can re-express $\hat{m}_t$ in terms of the mass of the top quark. This gives
\begin{equation}
\hat{m}_t = \frac{m_t}{\sin\alpha}\sqrt{\frac{(M_q^2-m_t^2)(M_q^2-\cos^2\alpha\ m_t^2)}
{\ M_q^4-2\csc^2\alpha\ m_t^2 M_q^2+2\cot^2\alpha\ m_t^4}} \ ,
\label{eq:yt}
\end{equation}
where $\alpha$ is still defined by Eq.~(\ref{eq:alpha}). This shows also that the inequality
\begin{equation}
\sin^2\alpha > \frac{2m_t^2(M_q^2-m_t^2)}{M_q^4-2m_t^4}
\label{eq:boundalpha}
\end{equation}
must be satisfied in order for the top quark mass to attain its observed value. Here the SM limit is achieved by taking 
both $\mu_q\to\infty$ and $M_q\to\infty$, with $\mu_q/M_q$ finite \cite{SekharChivukula:2009if}.

The bottom quark fields are diagonalized as in Eq.~(\ref{eq:bottom}), with the heavy bottom mass still given by Eq.~(\ref{eq:Bmass}). The top quark fields are diagonalized by $3\times 3$ matrices:
\begin{equation}
\left(\begin{array}{c} t_{0L} \\ t^q_{1L} \\ t^\Psi_{1L} \end{array}\right) = L_t
\left(\begin{array}{c} t_L \\ T^q_L \\ T^\Psi_L \end{array}\right) \ , \quad
\left(\begin{array}{c} t_{0R} \\ t^q_{1R} \\ t^\Psi_{1R} \end{array}\right) = R_t
\left(\begin{array}{c} t_R \\ T^q_R \\ T^\Psi_R \end{array}\right) \ .
\label{eq:diag1}
\end{equation}
The perturbative expressions (in powers of $m_t/M_q$) we obtained for the heavy top masses and the rotation matrices $L_t$ and $R_t$ are straightforward to calculate, but lengthy and not informative to look at.

\begin{figure}[!t]
\includegraphics[width=7cm,height=5cm]{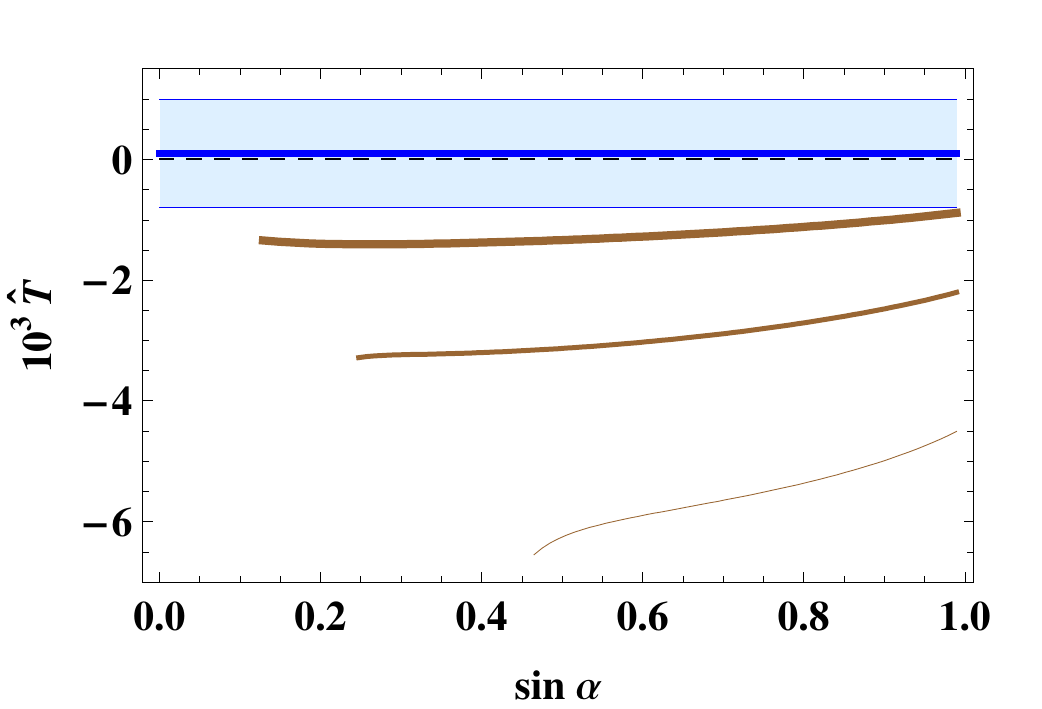}
\includegraphics[width=7cm,height=5cm]{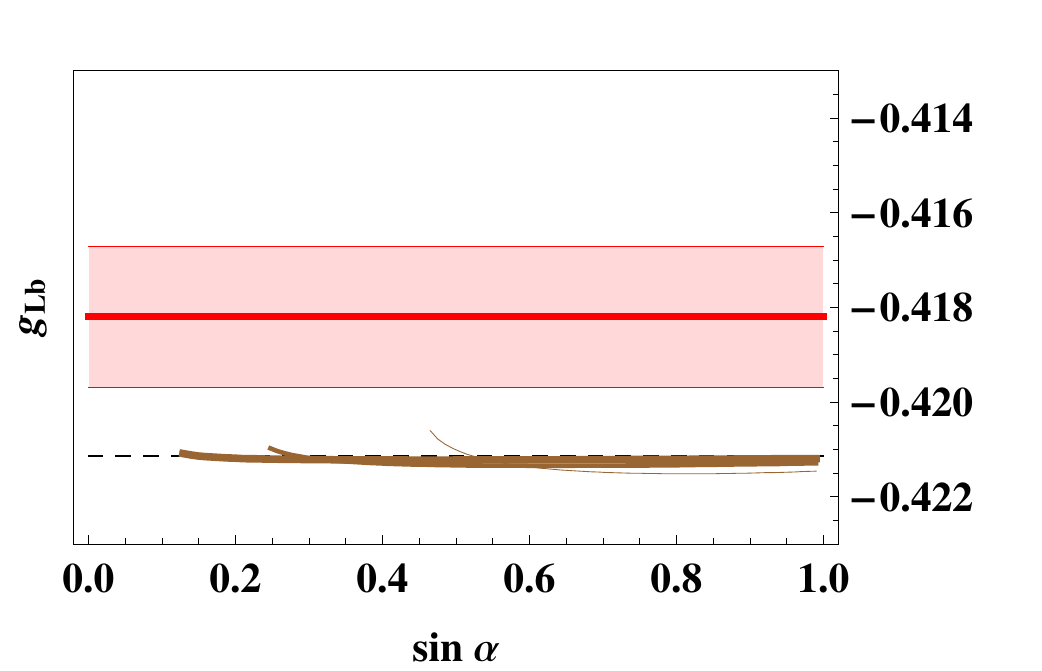}
\caption{$\hat{T}$ (left) and $\delta g_{Lb}$ (right) in the model of Eq.~(\ref{eq:Lbidoublet1}). The thin to thick curves are for $M_q=$ 0.5, 1, and 2 TeV, respectively, whereas the dashed lines are the SM predictions. In the plot for $\hat{T}$ the experimental $1\sigma$ band shown is for $m_H=115$ GeV; the band for larger $m_H$ lies at higher positive valuesof $\hat{T}$ \protect\cite{Barbieri:2004qk}. The experimental $1\sigma$ band for $\delta g_{Lb}$ has a negligible dependence on the Higgs mass \protect\cite{:2005ema}. The theoretical curves cut off at low values of $\sin\alpha$, as required by the bound from Eq.~(\ref{eq:boundalpha}). }
\label{fig:plotbidoublet}
\end{figure}

The one-loop contribution to $\hat{T}$ is given by the diagrams of Fig.~\ref{fig:diagrams}(a), where in gauge eigenstate basis $u_i=t_0,t_1^q,\Omega_1$ and $d_i=b_0,b_1,t_1^\Psi$. The correction to $g_{Lb}$ is given by the diagrams of Fig.\ref{fig:diagrams}(b), where, in mass eigenstate basis, $t_i=t,T^q,T^\Psi$. A straightforward calculation yields
\begin{eqnarray}
\hat{T}&=&\hat{T}^{\rm SM}\Bigg[-\frac{1+\cos^2\alpha}{\cos^2\alpha}\left(8\log\frac{M_q/\cos\alpha}{m_t}-\frac{22}{3}\right)
+\frac{4}{3}\ \frac{6-9\sin^2\alpha+5\sin^4\alpha-\sin^6\alpha}{\sin^8\alpha} \nonumber \\
&&\quad\quad\ \  -\frac{3+11\cos 2\alpha+\cos 4\alpha+\cos 6\alpha}{\sin^{10}\alpha\ \cos^2\alpha}\log\frac{1}{\cos\alpha}\Bigg]\frac{m_t^2}{M_q^2/\cos^2\alpha}
+{\cal O}\left(\frac{m_t^4}{M_q^4}\right) \label{eq:Thatbidoublet1} \\
\delta g_{Lb}&=&\delta g_{Lb}^{\rm SM}\Bigg[
\left(\frac{5}{8\sin^2\alpha}-\cos^2\alpha-1\right)\log\frac{M_q/\cos\alpha}{m_t}
-\frac{5}{8\sin^2\alpha}\log\frac{M_q}{m_t}+\frac{3}{8\sin^2\alpha}\log\frac{1}{\cos\alpha}\Bigg]\frac{m_t^2}{M_q^2}
+{\cal O}\left(\frac{m_t^4}{M_q^4}\right) \label{eq:gLbbidoublet1}
\end{eqnarray}
The contribution to $\hat{S}$ is
\begin{eqnarray}
\hat{S}=\frac{1}{96\pi^2}\left[-\frac{1+\cos^2\alpha}{\cos^2\alpha}\left(8\log\frac{M_q}{m_t}-15\right)
+8\cot^2\alpha\left(1+\log\frac{1}{\cos\alpha}\right)\right]\frac{m_t^2}{M_q^2/\cos^2\alpha}
\label{eq:Shatbidoublet}
\end{eqnarray}
Once again, the leading order contributions to $\hat{S}$ and $\hat{T}$ are both proportional to $m_t^2/M_q^2$, and $\hat{S}$ is numerically much smaller than $\hat{T}$.

In this case the largest contributions to $\delta g_{Lb}$ cancel, because of the custodial $O(4)$ symmetry; the remainder has a small and generally negative value. The dominant log contribution to $\hat{T}$ is clearly {\it negative}, with a rather large numerical coefficient. As illustrated in Fig. \ref{fig:plotbidoublet}, a model based strictly on the present Lagrangian, Eq.~(\ref{eq:Lbidoublet1}), does not agree with the data on its own; some other physics yielding positive corrections to both $\hat{T}$ and $g_{Lb}$ would be needed to restore agreement with data at the 1-sigma level.   In principle, the situation could be improved by adding a heavy vector singlet to the present model; this would not introduce sources of O(4)  hard breaking and, as demonstrated in Fig. \ref{fig:plotsinglet}, it could provide the needed positive corrections.  However, there is also a more economical alternative, which we explore in the next section of the paper.

\subsection{Bi-Doublet and Singlet}\label{sec:3B}

We have just seen that a heavy vector-like bi-doublet, mixed with the SM-like left-handed top-bottom weak doublet, gives a negative contribution to $\hat{T}$ and a very small contribution to $g_{Lb}$: this means that the predictions for $\hat{T}$ and $g_{Lb}$ are both below the $\pm 1\sigma$ experimental bands. On the other hand, in section~\ref{sec:2A} we observed that adding a vector singlet gives positive (and correlated) corrections to both $\hat{T}$ and $g_{Lb}$. Therefore the most obvious thing to do, in order to restore agreement with experiment, is to build a model with a vector bi-doublet $Q_1$ (mixed with the SM-like left-handed top-bottom doublet $q_{0L}$), a vector singlet $t_1$ (mixed with the SM-like right-handed top singlet $t_{0R}$), and an $O(4) $-symmetric Yukawa interaction. In order to simplify our analysis we will let the mixing mass between $q_{0L}$ and $q_{1R}$ become infinite: this effectively removes $q_{0L}$ and $q_{1R}$ from the Lagrangian, and promotes $q_{1L}$ to the role of SM-like left-handed top-bottom doublet. In Eq.~(\ref{eq:Lbidoublet1}) this is achieved by letting $\mu_q$ become infinite, whence $\sin\alpha\to 1$. We would like to stress, however, that this is only assumed for the sake of simplicity: allowing for a finite $\mu_q$ does not qualitatively modify our results.

 In that context, let us consider a model in which a SM-like left-handed top-bottom doublet is part of a bi-doublet $Q_{1L}$: 
\begin{eqnarray}
Q_{1L}\equiv\left(\begin{array}{cc} q_{1L} & \Psi_{1L} \end{array}\right)
=\left(\begin{array}{cc}t^q_{1L} & \Omega_{1L} \\ b_{1L} & t^\Psi_{1L} \end{array}\right) \ .
\end{eqnarray}
Unlike in the previously analyzed models, here the field playing the role of the SM-like fundamental top-bottom doublet is denoted $q_{1L}$ and not $q_{0L}$, as consistent with the factors discussed above. As in the model of section~\ref{sec:2A}, the SM-like right-handed singlet, $t_{0R}$, is allowed to mix with a vector singlet $t_1$. The structure of this model corresponds to the ``doublet extended" standard model
of \cite{SekharChivukula:2009if}, augmented by the additional $t_1$ weak/custodial singlet field.
Imposing an unbroken $O(4) $ symmetry in the Yukawa sector leads to the top-sector Lagrangian
\begin{eqnarray}
{\cal L}_{\rm top} &=& \bar{Q}_{1L} i \slashed{D} Q_{1L} 
+ \bar{t}_{0R} i \slashed{D} t_{0R}
+ \bar{t}_1 i \slashed{D} t_1
+ \bar{\Psi}_{1R} i \slashed{D} \Psi_{1R} \nonumber \\
&-&\mu_t\left(\bar{t}_{1L}t_{0R} + {\rm h.c.}\right)
-M_q \left(\bar{\Psi}_L\Psi_R + {\rm h.c.} \right)
-M_t \bar{t}_1 t_1
- y_t \left({\rm Tr}\ \bar{Q}_{1L}\Phi\ t_{1R} + {\rm h.c.}\right) \ .
\label{eq:Lbidoublet2}
\end{eqnarray}
Custodial $O(4) $ is now softly broken by $M_q$ and, as in the model of Eq.~(\ref{eq:Lbidoublet1}), we can safely ignore the small  $O(4) $ breaking due to gauge interactions. Here, the SM limit corresponds to taking $\mu_t\to\infty$ and $M_q\to\infty$; as in the previous section, the operator arising from integrating out the heavy states will have the form of ${\cal O}_2$ in  Eq.~(\ref{eq:extl-op}).

The mass Lagrangian is
\begin{equation}
{\cal L}_{\rm mass} = -
\left(\begin{array}{ccc} t^q_{1L} & t^\Psi_{1L} & t_{1L} \end{array}\right)
\left(\begin{array}{ccc} 0 & 0 & \hat{m}_t \\ 0 & M_q & \hat{m}_t \\ \mu_t & 0 & M_t \end{array}\right)
\left(\begin{array}{c} t_{0R} \\ t^\Psi_{1R} \\ t_{1R} \end{array}\right)
- M_q\bar{\Omega}_{1L}\Omega_{1R}+ {\rm h.c.} \ ,
\end{equation}
where $\hat{m}_t$ is still given by Eq.~(\ref{eq:mthat}). As usual, we can rewrite $\hat{m}_t$ in terms of the mass of the top quark. This gives  
\begin{equation}
\hat{m}_t = \frac{m_t}{\sin\beta}\sqrt{\frac{(M_q^2-m_t^2)(M_t^2-\cos^2\beta\ m_t^2)}{(M_q^2-2m_t^2)(M_t^2-\cot^2\beta\ m_t^2)}} \ ,
\label{eq:mthatmt}
\end{equation}
whence
\begin{equation}
M_q > \sqrt{2}\ m_t\ ,\quad \tan\beta > \frac{m_t}{M_t} \ .
\label{eq:boundbeta}
\end{equation}
The top quark fields are diagonalized by $3\times 3$ matrices:
\begin{equation}
\left(\begin{array}{c} t^q_{1L} \\ t^\Psi_{1L} \\ t_{1L} \end{array}\right) = L_t
\left(\begin{array}{c} t_L \\ T^\Psi_L \\ T_L \end{array}\right) \ , \quad
\left(\begin{array}{c} t_{0R} \\ t^\Psi_{1R} \\ t_{1R} \end{array}\right) = R_t
\left(\begin{array}{c} t_R \\ T^\Psi_R \\ T_R \end{array}\right) \ .
\end{equation} 
As in the heavy bi-doublet scenario of the last section, the perturbative expressions for masses and rotation matrices are straightforward but quite lengthy, and will not be shown in here.

\begin{figure}[!t]
\includegraphics[width=5cm,height=4.5cm]{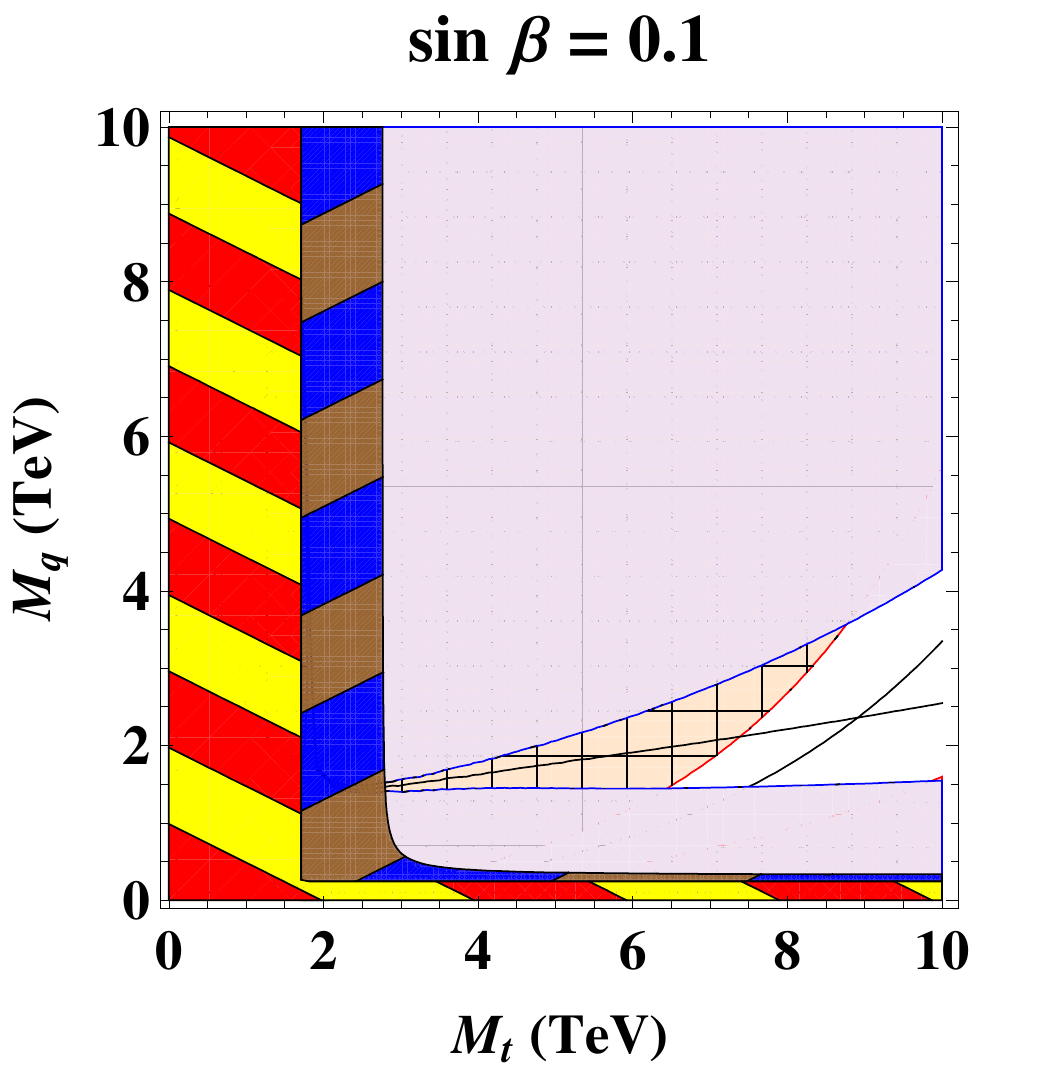}
\includegraphics[width=5cm,height=4.5cm]{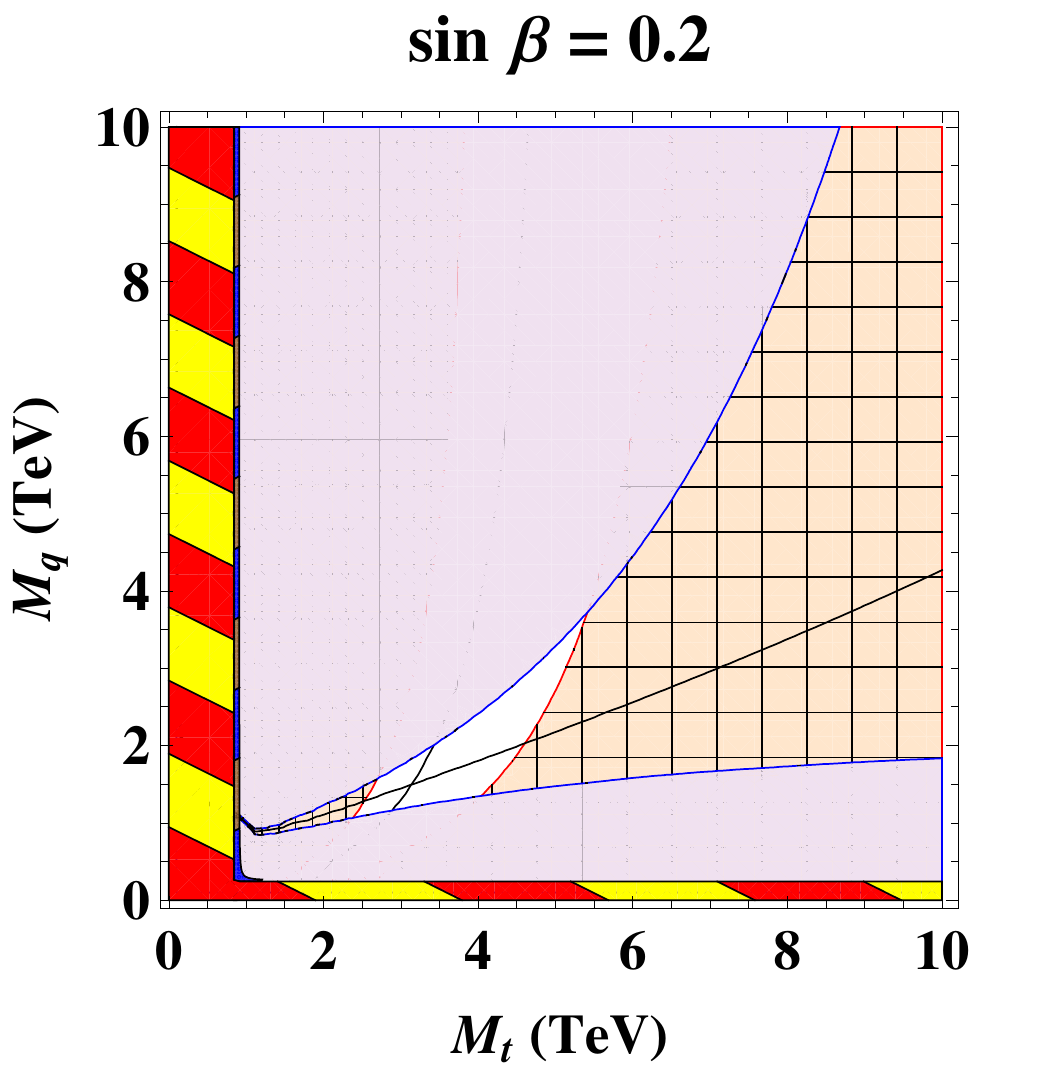}
\includegraphics[width=5cm,height=4.5cm]{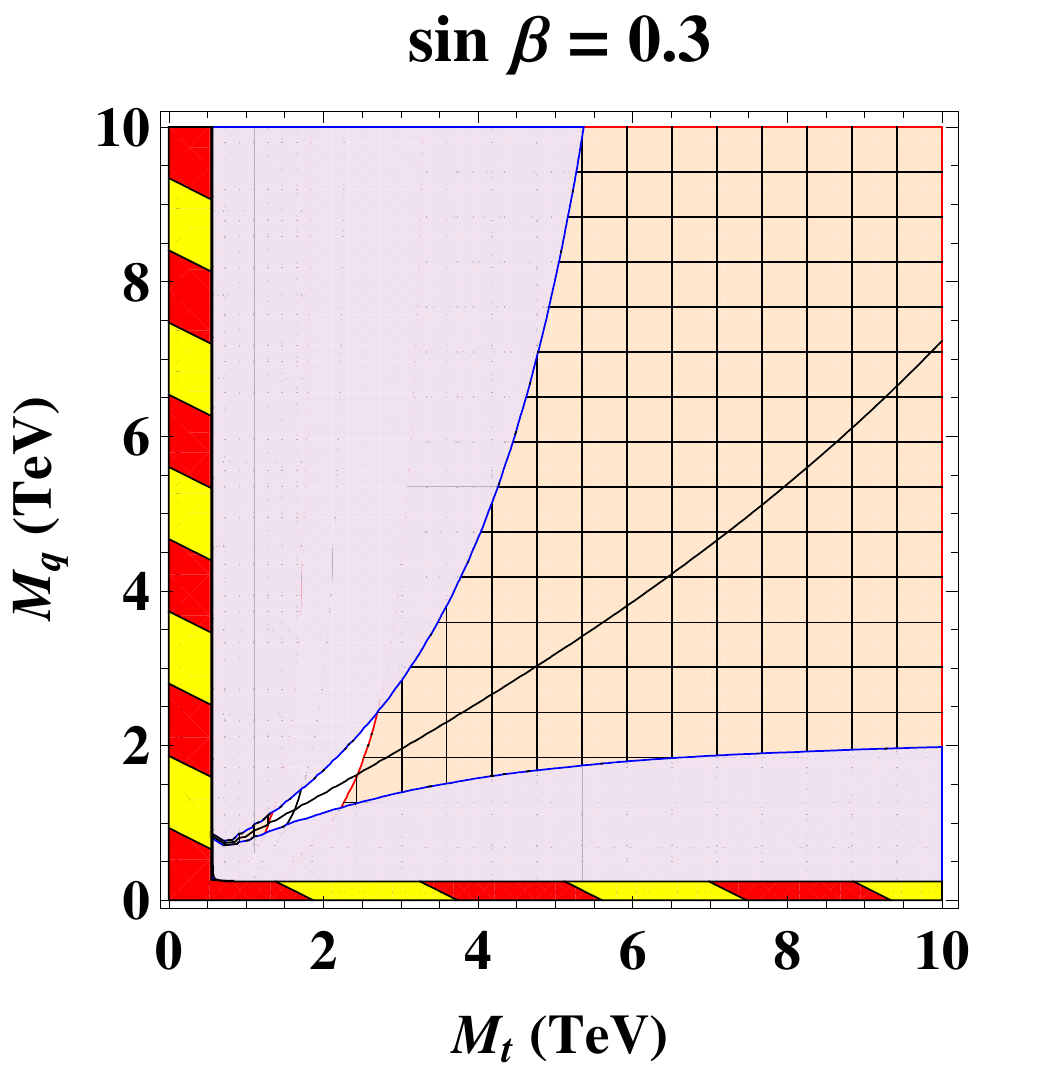} \\
\vspace{0.5cm}
\includegraphics[width=5cm,height=4.5cm]{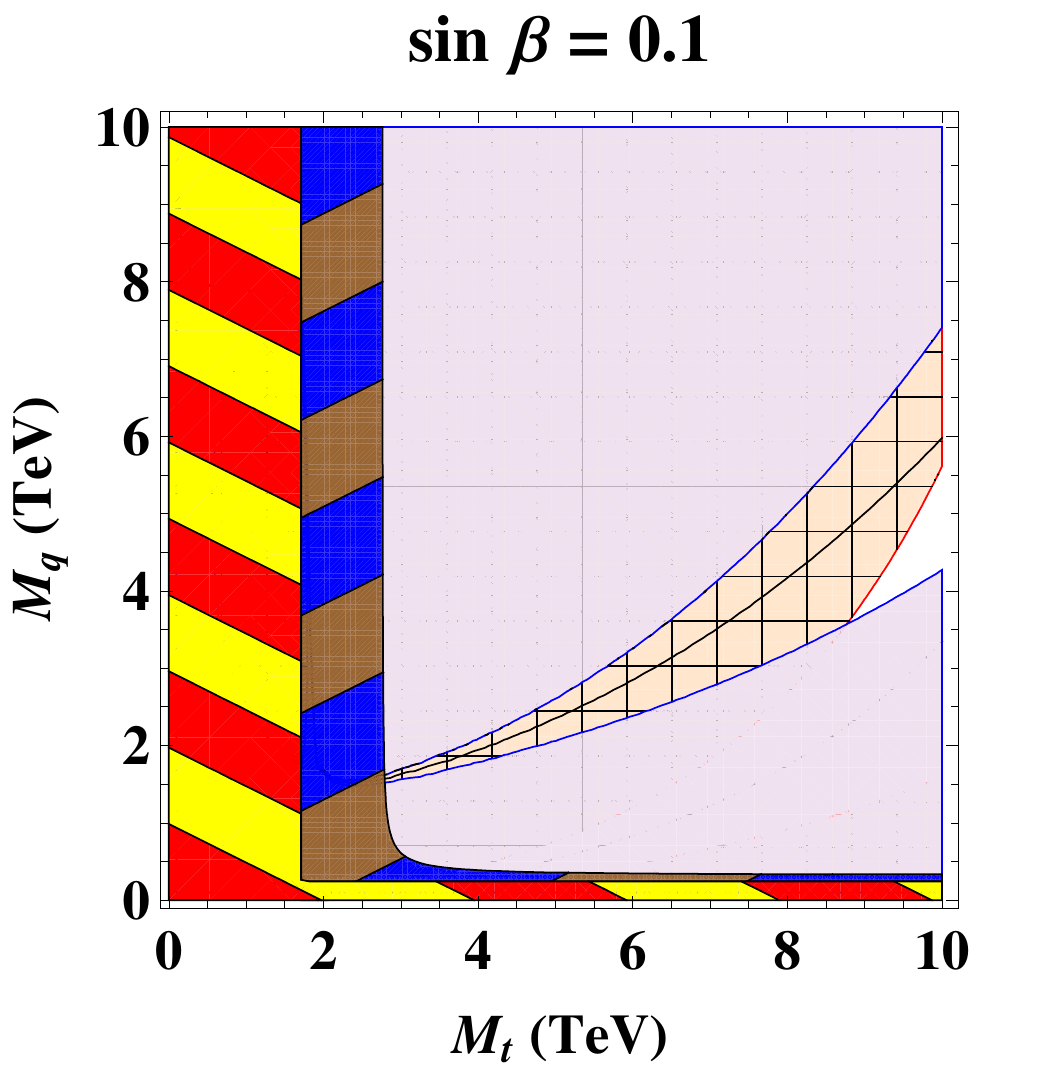}
\includegraphics[width=5cm,height=4.5cm]{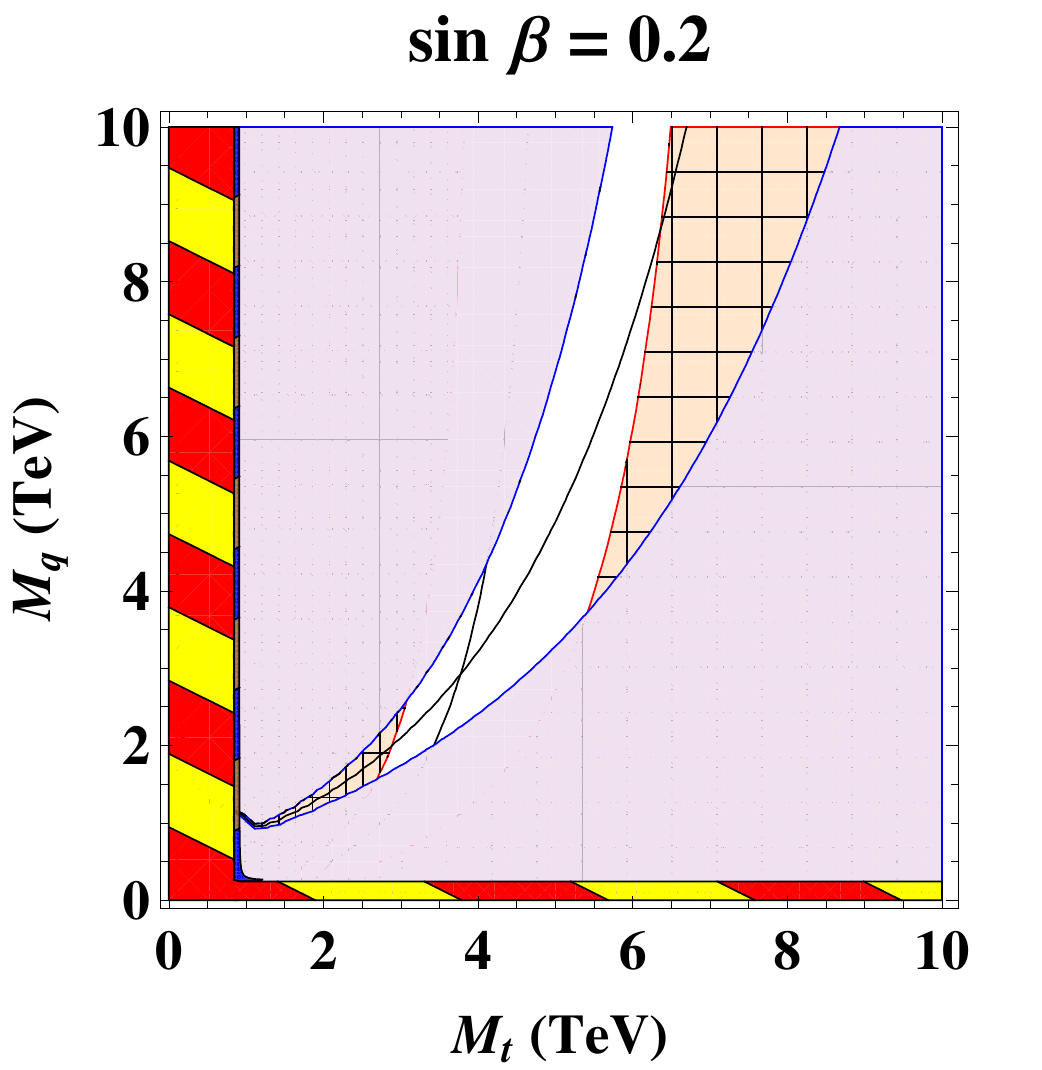}
\includegraphics[width=5cm,height=4.5cm]{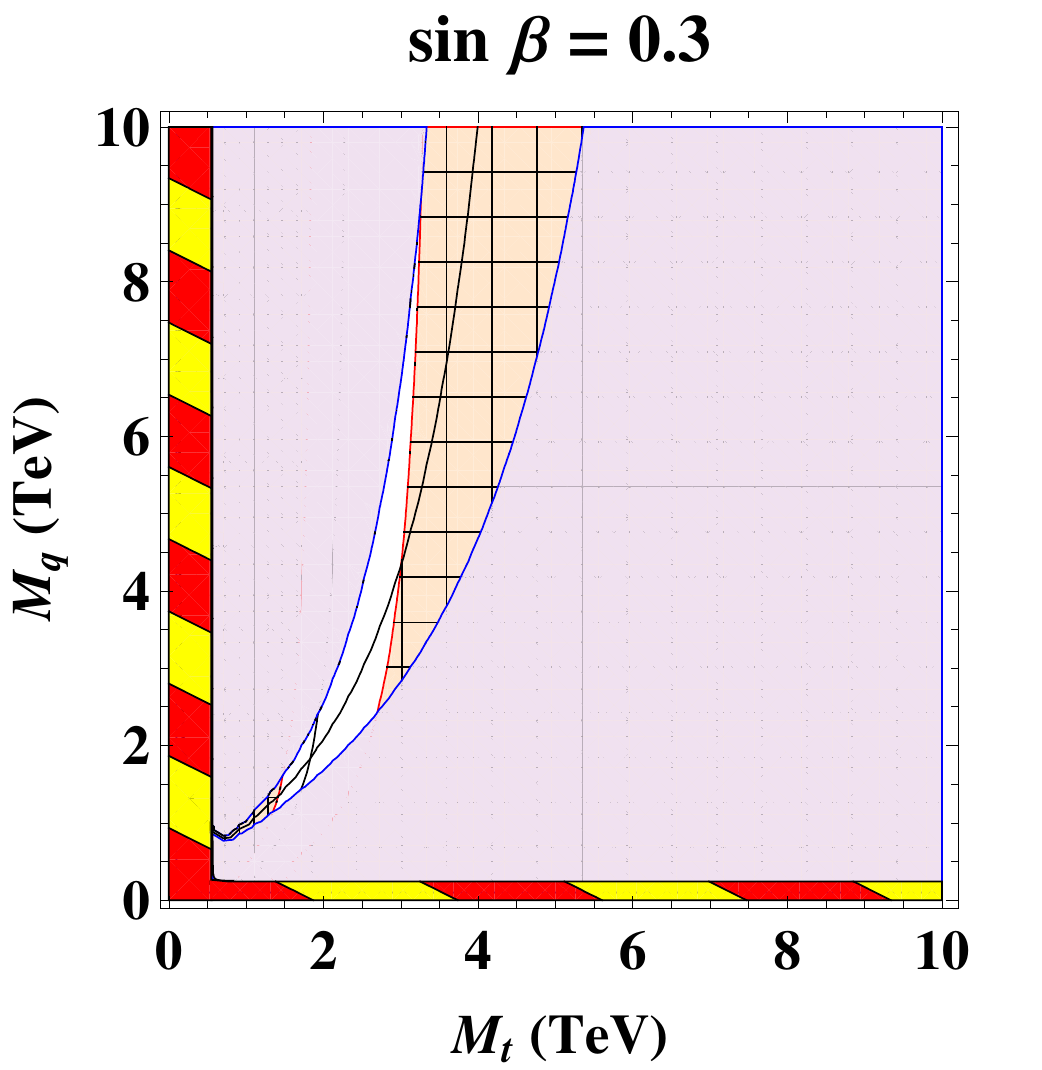}
\caption{Allowed and excluded regions for the bi-doublet plus singlet model described in Section \ref{sec:3B},  at the $1\sigma$ level, in the $M_t-M_q$ plane, for $\sin\beta=0.1,0.2,0.3$. The top row (bottom row) graphs are for $m_H=115\ (800)$ GeV. Only the white regions are allowed; see text for details about the various excluded (shaded, striped, hatched) zones.}
\label{fig:constraints}
\end{figure}

The one-loop contribution to $\hat{T}$ is given by the diagrams of Fig.~\ref{fig:diagrams}(a), where in the gauge eigenstate basis $u_i=t^q_1,\Omega_1$ and $d_i=b_1,t_1^\Psi$. The perturbative expressions are quite complicated, so we will only show the results for the limiting cases $M_q\to\infty$ and $M_t\to\infty$, respectively:
\begin{eqnarray}
\lim_{M_q\to\infty}\hat{T}&=&\hat{T}^{\rm SM} \left(4\log\frac{M_t/\cos\beta}{m_t}+\frac{1}{\sin^2\beta}-3\right)
\frac{\cos^4\beta\ m_t^2}{\sin^2\beta\ M_t^2}\left[1+{\cal O}\left(\frac{m_t^4}{M_t^4}\right)\right] \ , \nonumber \\
\lim_{M_t\to\infty}\hat{T}&=&\hat{T}^{\rm SM}\left(-8\log\frac{M_q}{m_t}+\frac{22}{3}\right)\frac{m_t^2}{M_q^2}
\left[1+{\cal O}\left(\frac{m_t^4}{M_q^4}\right)\right] \ .
\end{eqnarray}
The $M_q\to\infty$ limit is positive, and agrees with Eq.~(\ref{eq:singlet}). The $M_t\to\infty$ limit is negative, and agrees with Eq.~(\ref{eq:Thatbidoublet1}) for $\sin\alpha\to 1$. For the $\hat{S}$ parameter we obtain
\begin{eqnarray}
\lim_{M_q\to\infty}\hat{S}&=&\frac{g^2}{96\pi^2}\left(4\log\frac{M_t/\cos\beta}{m_t}-5\right)
\frac{\cos^4\beta\ m_t^2}{\sin^2\beta\ M_t^2}\left[1+{\cal O}\left(\frac{m_t^4}{M_t^4}\right)\right] \ , \nonumber \\
\lim_{M_t\to\infty}\hat{S}&=&-\frac{1}{96\pi^2}\left(8\log\frac{M_q}{m_t}-15\right)\frac{m_t^2}{M_q^2}
\left[1+{\cal O}\left(\frac{m_t^4}{M_q^4}\right)\right]\ ,
\end{eqnarray}
respectively in agreement with Eq.~(\ref{eq:Shatsinglet}) and Eq.~(\ref{eq:Shatbidoublet}) with $\sin\alpha\to 1$. Once again, $|\hat{S}/\hat{T}|\ll 1$.

The correction to $g_{Lb}$ is given by the diagrams of Fig.\ref{fig:diagrams}(b), where, in the mass eigenstate basis, $t_i=t,T^\Psi,T$. Since the perturbative expressions are complicated, we only show the $M_q\to\infty$ and $M_t\to\infty$ limits, respectively:
\begin{eqnarray}
\lim_{M_q\to\infty}\delta g_{Lb}&=&\delta g_{Lb}^{\rm SM} \left(4\log\frac{M_t/\cos\beta}{m_t}+\frac{1}{\sin^2\beta}-3\right)
\frac{\cos^4\beta\ m_t^2}{\sin^2\beta\ M_t^2}\left[1+{\cal O}\left(\frac{m_t^4}{M_t^4}\right)\right] \ , \nonumber \\
\lim_{M_t\to\infty}\delta g_{Lb}&=&\delta g_{Lb}^{\rm SM}\left(-\log\frac{M_q}{m_t}\right)\frac{m_t^2}{M_q^2}
\left[1+{\cal O}\left(\frac{m_t^4}{M_q^4}\right)\right] \ .
\end{eqnarray}
The $M_q\to\infty$ limit is positive, and agrees with Eq.~(\ref{eq:singlet}). The $M_t\to\infty$ limit is small and negative, and agrees with Eq.~(\ref{eq:gLbbidoublet1}) for $\sin\alpha\to 1$.

Since there are now two heavy fermion mass scales ($M_q$ associated with the bi-doublet and $M_t$ associated with the heavy singlet), we find convenient to plot the regions allowed by data on $\hat{T}$ and $g_{Lb}$ in the plane defined by these mass scales. In Fig.~\ref{fig:constraints} we show the allowed and excluded regions, at the $1\sigma$ level, in the $M_t-M_q$ plane, for $\sin\beta=0.1,0.2,0.3$. The top row graphs are for $m_H=115$ GeV, whereas the bottom row graphs are for $m_H=800$.  The L-shaped striped region covering the lower left-corner (thinner yellow-red stripes) is excluded by the need to reproduce the measured top mass, as required by Eq.~(\ref{eq:boundbeta}). In the adjacent striped region (thicker, blue-brown stripes) the top Yukawa coupling is larger than $4\pi$: this is mainly possible for small values of $\sin\beta$, as shown by Eq.~(\ref{eq:mthatmt}) and Eq.~(\ref{eq:mthat}). The purple monochrome shaded region is excluded by constraints on $\hat{T}$: in the upper left region $\hat{T}$ is large and positive, in the lower right $\hat{T}$ is large and negative, whereas the curve centered between the boundaries of the purple monochrome shaded regions traces the locus of the expectation value of $\hat{T}$. The yellow hatched region is excluded by constraints on $\delta g_{Lb}$: in the left region $\delta g_{Lb}$ is large and positive, in the right region $\delta g_{Lb}$ is very small, and the curve centered between the boundaries of the yellow hatched regions traces the locus of the expectation value of $\delta g_{Lb}$.  Finally, the white regions are consistent with all constraints.  We see that, for low values of $\sin\beta$, heavy vector fermions with masses of order 2 TeV or more are allowed.

Note also that if we, alternatively, assume that the model includes no new corrections to $g_{Rb}$ beyond those in the SM, then regions of Fig. 5 where $\delta g_{Lb} \approx 0$ would, instead, be acceptable at the 99\% CL (these regions lie within the yellow hatched spaces to the right of the white regions in each pane of Fig. 5).   Again, the heavy vector fermions would have masses of at least 2 TeV.

\section{Conclusions}\label{sec:conclusions}

In this paper we analyzed the effects of top-quark compositeness resulting from mixing between fundamental fields with SM-like quantum numbers and new composite vector fermions.  After reviewing the operators that are generated when heavy mostly-vector states are integrated out and identifying which would have the most significant impacts on low-energy observables, we constructed explicit models whose low-energy effective Lagrangians include those operators and studied the phenomenology in more detail.  Our analysis focused on the observables $\hat{T}$ and $g_{Lb}$, which are sensitive to operators that break the extended custodial $O(4) $ symmetry, while also commenting on the contributions of the new physics to $\hat{S}$. We considered two scenarios of $O(4) $ breaking: explicit breaking in the Yukawa interactions and soft breaking in mixing mass terms. Our models can be seen as highly deconstructed versions of extra-dimensional dual models featuring the same global symmetries.  In that language, we have studied the effects of the composite fermion sector only, and neglected the composite gauge sector. This is internally consistent, since the contribution from gauge Kaluza-Klein modes to $\hat{T}$ and $g_{Lb}$ is suppressed in models with custodial symmetry~\cite{Carena:2006bn}.

We found that models with explicit $O(4) $ breaking in the Yukawa sector are disfavored by experiment, as the contributions to $\hat{T}$ become too large in regions of the parameter space in which $g_{Lb}$ is within the $1\sigma$ bounds.  These models correspond to two of the scenarios that our effective field theory analysis had initially identified as potentially interesting.  When the right-handed top is composite and, in the limit $M_t \to \infty$ in Eq. (\ref{eq:Lsinglet}) the low-energy effective Lagrangian includes
 an operator of the ${\cal O}_3$ form $(\bar{q}_{0L}\tilde{\varphi}) i\slashed{D} (\tilde{\varphi}^\dagger q_{0L})$. On the other hand, when the left-handed top-bottom doublet is composite and, in the
limit $M_q\to \infty$ (with $\mu_q/M_q$ fixed) in Eq. (\ref{eq:Ldoublet}) one generates an operator of the ${\cal O}_1$ form
$(\bar{t}_{0R} \tilde{\varphi}^\dagger) i \slashed{D} (\tilde{\varphi} t_{0R})$. Unfortunately, the presence of either of these operators leads to correlated positive contributions to $\hat{T}$  and $\delta g_{Lb}$ that are inconsistent with the data if no other new physics is present.

 In models with soft $O(4) $ breaking, the left-handed top-bottom doublet is mixed with a vectorial quark bi-doublet of $SU(2)_L\times SU(2)_R$. The two $SU(2)_L$ doublets embedded in the bi-doublet give canceling contributions to $g_{Lb}$, and an overall negative contribution to $\hat{T}$: this agrees with the findings of Refs.~\cite{Carena:2006bn,Carena:2007ua}, where bulk fermion bi-doublets mix with the boundary-localized SM left-handed weak doublets. Adding an extra vectorial singlet provides positive and correlated corrections to $\hat{T}$ and $g_{Lb}$ that can improve the agreement between model and data: analogous results are found in extra-dimensional models with bulk singlets mixing with the brane-localized SM right-handed weak singlets~\cite{Carena:2006bn,Carena:2007ua}. Therefore, models of top-compositeness with soft $O(4) $ breaking feature a combination of positive and negative contributions to $\hat{T}$ and $g_{Lb}$, and agreement with experiment, at the $1\sigma$ level, is possible for both light- and heavy-Higgs regimes. Although our results agree with refs.~\cite{Carena:2006bn,Carena:2007ua}, our analysis further includes independent fermion mass scales, in addition to the arbitrary compositeness parameters $\sin\alpha$ and $\sin\beta$.

In the language of effective field theory \cite{Contino:2006nn,SekharChivukula:2009if}, we see that the patterns of deviation in $\hat{T}$ and
$\delta g_{Lb}$ that arise in the various models are strongly correlated with the quantum numbers of the heavy vector fermions with which
the top-quark mixes. In particular, the interesting and potentially {\it negative} contributions arise from the existence of the exotic
$2_{+7/6}$ state in the bidoublet field(s) introduced in models with extended $O(4)$ custodial symmetry \cite{Agashe:2006at}.  It is integrating out these
states that produces the crucial low-energy operator $(\bar{t}_{0R} \varphi^\dagger) i \slashed{D} (\varphi t_{0R})$ of the ${\cal O}_2$ form.
While introducing only these extra bi-doublet states does not give rise to a phenomenologically
acceptable model  \cite{SekharChivukula:2009if},
allowing the top quark to mix both with an
exotic $2_{+7/6}$ state, as well as an ordinary $1_{+2/3}$ singlet or $2_{+1/6}$ doublet state  (in section \ref{sec:3B}) can yield a phenomenologically viable model.
Such scenarios also necessarily include ``ditop" quarks ($\Omega$) with charge $+5/3$ which would decay into two like-sign $W$'s and a bottom quark and could provide a potentially interesting pair-production signature at the LHC \cite{Contino:2008hi}.

Finally, we note that there is another possible implementation of a a softly-broken extended $O(4)$ symmetry, in which the $t_R$ field (or the heavy field with which it mixes) is embedded in a  $(3,1)_{2/3} + (1,3)_{2/3}$ {\it triplet} field  \cite{Agashe:2006at}.  As shown in Appendix \ref{sec:appx}, such models generically include dangerous tree-level corrections to $\delta g_{Lb}$.  In the $O(4)$ limit where these tree-level corrections are absent, integrating out the heavy vector fermions gives rise to a low-energy operator of  {\it precisely} the same form
as ${\cal O}_3$.  The low-energy phenomenology of this model, therefore, is similar to those discussed above.

\begin{acknowledgments}
RSC and EHS were supported, in part, by the US National Science Foundation under grant PHY-0854889.  They acknowledge the hospitality of the Institute for Advanced Study (Princeton) and the Aspen Center for Physics, where part of this work was completed. The work of RF is supported by IISN and the Belgian Science Policy (IAP-VI-11).

\end{acknowledgments}

\appendix
\section{}
\label{sec:appx}

In the models studied in this paper, we have embeded the $t_R$ field (or the heavy fields
with which it mixes) in a $(1,1)_{2/3}$ custodial singlet field. Alternatively  \cite{Agashe:2006at},
one can choose to embed $t_R$ as a member of a $(3,1)_{2/3} + (1,3)_{2/3}$ {\it triplet}
field. Here, we construct the simplest realization of this possibility -- {\it i.e.} we construct
a model with a softly-broken extended custodial symmetry in which the $t_R$ is embedded
as a triplet.  While models including triplets generically yield tree-level contributions to the $Zb\bar{b}$ coupling, 
we show that these can be avoided in the $O(4)$-symmetric limit.  Finally, we discuss the form of the low-energy operators that arise
once soft symmetry breaking is introduced, and show that the operator is the familiar ${\cal O}_3$, whose phenomenology is studied elsewhere in this paper.

Let us consider an analog of the ``Doublet Extended Symmetry Breaking" model (DESM) of
\cite{SekharChivukula:2009if}. The Yukawa sector of such a model contains the following
terms,
\begin{equation}
\lambda_T {\rm Tr} (\bar{\Psi}_L \Phi \sigma^a) T^a_R + \lambda_F {\rm Tr}
(\bar{\Psi}_L \sigma^a \Phi) F^a_R~,
\end{equation}
where $\Psi_L=(q_L\  \chi_L)$ is a bidoublet fermion field (which transforms as a $(2,2)$ under
$O(4) \simeq SU(2) \times SU(2) \times P_{LR}$), $\Phi = (\tilde{\varphi}\ \varphi)$ is the usual bidoublet
Higgs field with $\tilde{\varphi} = i \sigma^2 \varphi^*$, $T^a$ and $F^a$ are $(1,3)$ and $(3,1)$ triplet
fermion fields, and the $\sigma^a$ are the usual Pauli matrices. 
The $P_{LR}$ symmetry requires that $\lambda_T=\lambda_F$.
We gauge the usual $SU(2) \times U(1)$ subgroup of $O(4) \times U(1)_X$, and
obtaining the correct fermion electric charges requires the $X$ charges of $\Psi_L$, 
$T_R$, and $F_R$ to be $+2/3$.

We may break the $O(4) \times P_{LR}$ symmetry of the top sector softly in the usual
way: we will include mass terms for all the extra fermions, just as in \cite{SekharChivukula:2009if}.
Hence, we will include chiral partners and fermion mass terms for the $\chi_L$ and $F_R$ fields.
The $T_R$ field includes two exotic fermions and, as its $a=3$ member, the $t_R$. Following
\cite{SekharChivukula:2009if}, we will add chiral partners and fermions masses for
the $T^{1,2}_R$ fields -- but leave $T^3_R = t_R$ massless and without a chiral partner.
The model now has precisely the same properties as the DESB model -- the top yukawa sector
has an extended $O(4) \times P_{LR}$ symmetry which is softly broken by the mass terms
for the extra fermions, and it reduces to the standard model in the limit where the extra fermion masses become large.

\subsection{Low-Energy Effects}

We now consider the form of the low-energy operators induced after integrating out
the heavy triplet fermion states. We will do so in two steps: first, let us integrate out complete
sets of fermion triplets $T^a$ and $F^a$, with masses $M_T$ and $M_F$. As we noted
above, however, we should not integrate out the state $T^3_R = t_R$ -- so we will add back in the
contribution of this state and see what kind of operator results.

\subsubsection{Complete Right-Handed Multiplets}

Let us begin by integrating out all six states in the complete right-handed
multiplet. Doing so results in an operator of
the following form
\begin{align}
\frac{\lambda^2_T}{M^2_T} 
\left[{\rm Tr}(\bar{\Psi}_L \Phi \sigma^a)i\fsl{D}(\sigma^a \Phi^\dagger \Psi_L)\right]
+ \frac{\lambda^2_F}{M^2_F} 
\left[{\rm Tr}(\bar{\Psi}_L \sigma^a \Phi )i\fsl{D}(\Phi^\dagger \sigma^a \Psi_L)\right]~.
\label{eq:opTF}
\end{align}
We proceed by using the Fierz identity\footnote{Recall that
$\sigma^a/2$ are the conventionally normalized $SU(2)$ generators.}
\begin{equation}
\sigma^a_{ij} \sigma^a_{kl} = 2 \delta_{il} \delta_{kj} - \delta_{ij}\delta_{kl}~.
\end{equation}
The first term in the operator then becomes
\begin{equation}
\frac{\lambda^2_T}{M^2_T} 
\left[2\,{\rm Tr}\left(\bar{\Psi}_L \Phi i\fsl{D} (\Phi^\dagger \Psi_L)\right)
- ({\rm Tr} \bar{\Psi}_L \Phi) i \fsl{D} ({\rm Tr} \Phi^\dagger \Psi_L)\right]~.
\label{eq:opT}
\end{equation}
In the low energy Lagrangian, we make the substitution
\begin{equation}
\Psi \to
\begin{pmatrix}
t_L & 0 \\
b_L & 0
\end{pmatrix}~,
\label{eq:sub}
\end{equation}
since the heavy exotic $\chi$ fields are decoupled. Therefore
\begin{equation}
{\rm Tr}(\Phi^\dagger \Psi_L) \to \tilde{\varphi}^\dagger q_L~,
\label{eq:singlet-t}
\end{equation}
and the second term in the operator yields precisely the familiar
${\cal O}_3 = (\bar{q}_L \tilde{\varphi}) i\fsl{D} (\tilde{\varphi}^\dagger q_L)$ of Eq. (\ref{eq:smtr-op}). 
In particular, this combination does not affect the $Z b\bar{b}$ vertex at tree-level.

Using the Leibniz product rule, the first term in Eq. (\ref{eq:opT}) can be manipulated
as follows
\begin{equation}
{\rm Tr}\left[\bar{\Psi}_L \Phi i\fsl{D} (\Phi^\dagger \Psi_L)\right] = 
{\rm Tr}\left[\bar{\Psi}_L \Phi \Phi^\dagger (i\fsl{D} \Psi_L)\right]
+ {\rm Tr} \left[\bar{\Psi}_L (\Phi i \fsl{D} \Phi^\dagger) \Psi_L \right]~,
\end{equation}
After electroweak symmetry breaking, $\langle \Phi \rangle = v^2 {\cal I}/2 \neq 0$, the first
contribution yields a renormalization of the usual $q_L$ kinetic energy term and produces no observable
effects. The second contribution, however, renormalizes the $Z$ and $W$ couplings of $q_L$. Using
\begin{equation}
D_\mu \Phi^\dagger = \partial_\mu \Phi^\dagger - i g W^a_\mu \Phi^\dagger\frac{\sigma^a}{2}
+ i g' B_\mu \frac{\sigma^3}{2} \Phi^\dagger~,
\end{equation}
we see that
\begin{equation}
\langle \Phi  i D_\mu  \Phi^\dagger \rangle \to
\frac{e \,v^2}{2 \sin\theta_W} \begin{pmatrix}
\frac{Z_\mu}{2 \cos \theta_W} & \frac{W^+_\mu}{\sqrt{2}} \\
\frac{W^-_\mu}{\sqrt{2}} & -\,\frac{Z_\mu}{2 \cos \theta_W}
\end{pmatrix}~.
\label{eq:EW}
\end{equation}
Defining $\delta g_{Lb}$ in terms of the $Zb_L\bar{b}_L$ coupling
\begin{equation}
\frac{e}{\sin\theta_W \cos\theta_W} \left(-\,\frac{1}{2} +\delta g_{Lb} + \frac{1}{3} \sin^2\theta_W\right)~,
\end{equation}
we therefore find that this second contribution from $T_R$ exchange yields
\begin{equation}
\delta g^T_{Lb} = -\, \frac{\lambda^2_T v^2}{2 M^2_T}~.
\end{equation}
The analysis of the $F^a$ exchange contributions in Eq. (\ref{eq:opTF}) proceeds
in a similar fashion, and we find 
\begin{equation}
\delta g^F_{Lb} = +\, \frac{\lambda^2_F v^2}{2 M^2_F}~.
\end{equation}

We therefore find that the tree-level shift in the $Zb\bar{b}$ coupling
that arises from integrating out a complete right-handed custodial triplet multiplet
(as summarized in the operator in Eq. (\ref{eq:opTF})) is
\begin{equation}
\delta g_{Lb} = \frac{v^2}{2}
\left(\frac{\lambda^2_F }{ M^2_F}
-\frac{\lambda^2_T }{ M^2_T}
\right)~.
\end{equation}
We explicitly see that, in the $O(4)$ limit where $\lambda_T=\lambda_F$
and $M_T=M_F$, there is no tree-level correction to the $Zb\bar{b}$ vertex.

\subsubsection{What about $t_R$?}

The calculation above integrated out the entire heavy $T^a$ multiplet, including the $T^3$ 
fermion. In our softly-broken symmetry model, however, $T^3_R = t_R$ remains
massless and has no chiral partner.  Hence, to obtain the net effect on the $Zbb$ vertex in our model, we
must isolate the specific contribution from the $T^3_R$ state to the $Zbb$ vertex.  
The  $T^3_R$ state gives rise to the following term in  Eq. (\ref{eq:opTF})
\begin{equation}
\frac{\lambda^2_T}{M^2_T} 
\left[{\rm Tr}(\bar{\Psi}_L \Phi \sigma^3)i\fsl{D}(\sigma^3 \Phi^\dagger \Psi_L)\right]~.
\label{eq:op3}
\end{equation}
Using the low-energy restriction of Eq. (\ref{eq:sub}), we see that
\begin{equation}
{\rm Tr}(\sigma^3 \Phi^\dagger \Psi_L) \to \tilde{\varphi}^\dagger q_L~,
\end{equation}
and hence the operator in Eq. (\ref{eq:op3}) just yields ${\cal O}_3$, an operator
we have already studied, and which does not affect the $Zb\bar{b}$ vertex at tree level.

\subsection{Conclusions}

In a model with a softly broken custodial symmetry with right-handed triplets, generically
$m_T \neq m_F$ and there can be dangerous tree-level corrections to $\delta g_{Lb}$.
When $m_T = m_F$, however, the effective field theory analysis reduces to the analysis
of operator ${\cal{O}}_3$ that is discussed in the main body of the paper.

\end{document}